\documentclass[12pt, a4paper]{article}
\usepackage[latin1]{inputenc}
\usepackage{vmargin}
\usepackage{amsfonts}
\usepackage{amssymb}
\usepackage{graphicx}
\usepackage{amsmath}
\usepackage{amsthm}
\usepackage{setspace}
\usepackage{multirow}
\usepackage{mathtools}
\usepackage{relsize}
\usepackage{cases}
\usepackage[bottom]{footmisc}
\setmarginsrb{1.5cm}{1.5cm}{1.5cm}{1.5cm}{0cm}{0cm}{0cm}{0.5cm}
\begin{document}

\newtheorem{Proposition}{Proposition}[section]
\newtheorem{Theorem}{Theorem}[section]
\newtheorem*{thma}{Theorem A.1}
\newtheorem*{thmb}{Theorem B.1}
\newtheorem{Corollary}{Corollary}[section]
\newtheorem{Definition}{Definition}[section]
\newtheorem{Lemma}{Lemma}[section]

\title{General Intensity Shapes in Optimal Liquidation\addtocounter{footnote}{-1}\thanks{This research was partially supported by the Research Initiative ``Microstructure des marchés financiers'' under the aegis of the Europlace Institute of Finance. The authors wish to acknowledge the helpful conversations with Yves Achdou (Université Paris-Diderot), Guy Barles (Université de Tours), Erhan Bayraktar (University of Michigan), Bruno Bouchard (Université Paris-Dauphine), Jean-Michel Lasry (Université Paris-Dauphine), Mike Ludkovski (UC Santa Barbara) and Nizar Touzi (Ecole Polytechnique). The anonymous referees also need to be warmly thanked for their thorough reading and the numerous improvements following their remarks.}}
\author{Olivier Guéant\addtocounter{footnote}{0}\thanks{UFR de Math\'ematiques, Laboratoire Jacques-Louis Lions, Universit\'e Paris-Diderot. 175, rue du Chevaleret, 75013 Paris, France. \texttt{olivier.gueant@ann.jussieu.fr}}, Charles-Albert Lehalle\addtocounter{footnote}{5}\thanks{Head of Quantitative Research. Cr\'edit Agricole Cheuvreux. 9, Quai du Pr\'esident Paul Doumer, 92400 Courbevoie, France. \texttt{clehalle@cheuvreux.com}} }
\date{}
\maketitle

\abstract{The classical literature on optimal liquidation, rooted in Almgren-Chriss models, tackles the optimal liquidation problem using a trade-off between market impact and price risk. It answers the general question of optimal scheduling but the very question of the actual way to proceed with liquidation is rarely dealt with. Our model, that incorporates both price risk and non-execution risk, is an attempt to tackle this question using limit orders. The very general framework we propose to model liquidation with limit orders generalizes existing ones in two ways. We consider a risk-averse agent whereas the model of Bayraktar and Ludkovski \cite{bayraktar2011liquidation} only tackles the case of a risk-neutral one. We consider very general functional forms for the execution process intensity, whereas \cite{GLFT} is restricted to exponential intensity. Eventually, we link the execution cost function of Almgren-Chriss models to the intensity function in our model, providing then a way to see Almgren-Chriss models as a limit of ours.\\}


\section{Introduction}

Since the late nineties and the first papers on the impact of execution costs on trading strategies (\emph{e.g.} \cite{bertsimas1998optimal}), an important literature has developed to tackle the problem of optimal liquidation. This literature, often rooted in the seminal papers by Almgren and Chriss \cite{almgren1999value,almgren2001optimal}, has long been characterized by a trade-off between, on the one hand, market impact that encourages to trade slowly and, on the other hand, price risk that provides an incentive to trade fast.\\

The first family of models, following Almgren and Chriss, considered general instantaneous price impact (sometimes called execution cost) and linear permanent price impact. Several generalizations have been proposed such as an extension to random execution costs \cite{almgren2003optimal}, or stochastic volatility and stochastic liquidity \cite{almgren2009optimal,almgren2011optimal}. Also, many objective functions to design optimal strategies have been proposed and discussed in order to understand the assumptions under which optimal strategies are deterministic (as opposed to adaptive). The initial mean-variance framework has been expressed in an expected utility setting (using a CARA utility function) in \cite{schied2010optimal} and \cite{gueantnew}, a mean-quadratic-variation framework has been considered \cite{forsyth2009optimal,tse2011comparison}, an initial-time mean-variance criterion has been discussed \cite{almgren2007adaptive,lorenz2010mean}, and the very interesting case of a general utility function has recently been considered in \cite{schied2009risk} to justify aggressive-in-the-money or passive-in-the-money strategies. Slightly different approaches have been proposed in this first generation of models (see \emph{e.g.} \cite{he2005dynamic} and \cite{huberman2005optimal}). They all derive from the initial models by Almgren and Chriss since market impact is either permanent or instantaneous. In other words, they do not take into account explicitly the resilience of the underlying order book.\\

Another family of models appeared following a paper by Obizhaeva and Wang \cite{obizhaeva2005optimal}. In these models, the limit order book is directly modeled and the authors consider its resilient dynamic after each trade. This second generation of optimal liquidation models, based on transient market impact, has developed in recent years (\cite{alfonsi2008constrained}, \cite{alfonsi2010optimal}, \cite{alfonsi2010optimal1} and \cite{predoiu2010optimal}). It raises the theoretical question of the functional forms for the transient market impact that are compatible with the absence of price manipulation (see \cite{alfonsi2009order}, \cite{gatheral2010no} and \cite{gatheral2010transient}).\\

All these models only make use of market orders, and hence only consider liquidity-taking strategies. They do not consider the possible use of limit orders that provide liquidity, nor the possible use of dark pools. Notwithstanding the preceding criticism, models \emph{à la} Almgren-Chriss provide a rather acceptable answer to the macroscopic question of the optimal scheduling of liquidation -- at least once the instantaneous market impact function has been replaced by an execution cost function modeling the ability to trade over short periods of time, with all possible means including limit orders, dark pools and market orders. However, they do not answer the question of the optimal way to proceed in practice and the methods currently used in the industry are seldom based on optimal control models at the microscopic level. This paper provides such a model of optimal liquidation using limit orders, and can be used, either to liquidate a portfolio as a whole over a few hours, or on shorter periods of time to follow a trading curve, be it a TWAP curve, a VWAP curve or an Almgren-Chriss (Implementation Shortfall) trading curve.\\

In our approach, a trader posts limit orders (thus providing liquidity instead of taking it) and does not know when his orders are going to be executed, if at all. As a consequence, the classical trade-off between market impact / execution cost and price risk is not central in our model. In our setting, a new risk is borne by the trader because execution is now a random process. This non-execution risk is very different, in its nature, from price risk. This new risk characterizes the recent literature on optimal liquidation, which focuses on the optimal way to liquidate rather than on optimal scheduling. The recent literature on optimal liquidation focuses indeed on alternatives to the use of market orders. Kratz and Schoneborn \cite{kratz2009optimal} proposed an approach inspired from models of the first family, but with both market orders and access to dark pools. Although they did not consider risk aversion with respect to the new risk borne by the trader, their model is one of the first in this new family of models. The optimal split of large orders across liquidity pools has then been studied by Laruelle, Lehalle and Pagès in \cite{laruelle2009optimal}. Liquidation with limit orders has been developed by Bayraktar and Ludkovski \cite{bayraktar2011liquidation} for general intensity functions but only in a risk-neutral framework. Guéant, Lehalle and Fernandez-Tapia \cite{GLFT} considered in parallel the specific case of an exponential intensity for a risk-averse agent. More recently, Huitema \cite{huitema2012optimal} considered liquidation involving market orders and limit orders, and Guilbaud and Pham \cite{guilbaud2012optimal} also proposed a liquidation model in a pro-rata microstructure.\\
One should also note that many models dealing with high-frequency market making have been developed that can be adapted to deal with optimal liquidation. Building on the model proposed by Ho and Stoll \cite{HoStoll} and then modified by Avellaneda and Stoikov \cite{avst08},\footnote{See \cite{citeulike:9272221} for the solution of the Avellaneda-Stoikov equations.} Cartea, Jaimungal and Ricci \cite{cartea2011buy} considered a model with exponential intensity, market impact on the limit order book, adverse selection effects and predictable $\alpha$. Cartea and Jaimungal \cite{cartea2012risk} recently used a similar model to introduce risk measures for high-frequency trading. Earlier, the same authors proposed a model \cite{cartea2010modeling} in which the reference price is modeled by a Hidden Markov Model. Eventually, Guilbaud and Pham \cite{guilbaud2011optimal} also used a model including both market orders and limit orders at best (and next to best) bid and ask together with stochastic spreads. As it is shown in appendix B, our model can be used to model trading on both sides of the market. Our choice to focus on optimal liquidation is mainly justified by practitioners' needs.\\

In this paper, we generalize both \cite{bayraktar2011liquidation} and \cite{GLFT}. We indeed consider both general shapes for the intensity functions, and an investor with a CARA utility function. Moreover, we present a limiting case in which the size of the orders tends to $0$ and we show that this limiting case is intrinsically linked to the usual continuous framework of Almgren and Chriss, although the latter framework only considers market orders. This limiting case helps to understand the meaning of intensity functions for quotes corresponding to marketable limit orders.\\

In Section 2, we present the setting of the model and the main hypotheses on execution. The third section is devoted to solving the partial differential equations arising from the control problem. Then, in Section 4, we provide illustrations of the model and we exhibit the asymptotic behavior of the quotes, generalizing therefore a result presented in \cite{GLFT}. Section 5 is dedicated to the study of a limit regime that corresponds to orders of small size. This fifth section leads to results linked to those obtained for the \emph{fluid limit} in \cite{bayraktar2011liquidation}, here in a risk-averse setting. This result is exploited in Section 6 that draws parallels between our model and the usual Almgren-Chriss framework.\\

\section{Optimal execution with limit orders: the model}

\subsection{Setup of the model}

Let us fix a probability space $(\Omega, \mathcal{F}, \mathbb{P})$ equipped with a filtration $(\mathcal{F}_t)_{t\geq 0}$ satisfying
the usual conditions. We assume that all random variables and stochastic processes are defined on $(\Omega, \mathcal{F},(\mathcal{F}_t)_{t\geq 0}, \mathbb{P})$.\\

We consider a trader who has to liquidate a portfolio containing a quantity $q_0>0$ of a given stock. We suppose that the reference price of the stock (that can be considered the first bid quote for example) follows a Brownian motion with a drift:
$$dS_t = \mu dt + \sigma dW_t.$$

To model limit orders and the execution process, we first introduce the set of admissible strategies:
$$\mathcal{A} = \left\lbrace (\delta_t)_{t \in [0,T]} | (\delta_t)_t \mathrm{\;predictable\;process\;}, \delta^- \in L^\infty(\Omega\times[0,T])\right\rbrace.$$

The trader under consideration continuously proposes an ask quote $S_t^a = S_t + \delta_t$. He will sell shares according to the rate of arrival of liquidity-taking orders at the price he quotes.\\

His inventory $q^\delta$, that is the number of shares he holds, evolves according to the following dynamics:
$$dq^\delta_t = - \Delta dN^{\delta}_t,$$ where $N^{\delta}$ is a point process giving the number of executed orders, each order being of size $\Delta$ -- we suppose that $\Delta$ is a fraction of $q_0$. The intensity process $(\lambda_t)_t$ of the point process $N^\delta$, that is the arrival rate of liquidity-taking orders, depends on both the (ask) price quoted by the trader and the size of its orders:

$$\lambda_t = \Lambda_{\Delta}(S^a_t-S_t)1_{q^\delta_{t-} > 0} = \Lambda_{\Delta}(\delta_t)1_{q^\delta_{t-} > 0},$$
where $\Lambda_{\Delta} : \mathbb{R} \to \mathbb{R}_+$ satisfies the following assumptions:

\begin{itemize}
\item $\Lambda_{\Delta}$ is strictly decreasing -- the cheaper the order price, the faster it will be executed,
\item $\lim_{\delta\rightarrow +\infty} \Lambda_{\Delta}(\delta) = 0$,
\item $\Lambda_{\Delta} \in C^2(\mathbb R)$,
\item $\Lambda_{\Delta}(\delta) \Lambda''_{\Delta}(\delta) \leq 2{\Lambda'}_{\Delta}(\delta)^2$.
\end{itemize}

As a consequence of his trades, the trader's cash account $X^{\delta}$ has the following dynamics:\\
$$dX^\delta_t = (S_t + \delta_t) \Delta dN^\delta_t.$$

Now, coming to the liquidation problem, the trader has a time horizon $T$ to liquidate his shares and his goal is to optimize the expected utility of his P\&L at time $T$. We focus on CARA utility functions so that the trader considers the following optimization problem:

$$\sup_{\delta \in \mathcal{A}} \mathbb{E}\left[- \exp\left(-\gamma\left(X^{\delta}_T+q^{\delta}_T (S_T-\ell(q^{\delta}_T))\right)\right) \right],$$
where $\gamma > 0$ is the absolute risk aversion parameter characterizing the trader, where $X^\delta_T$ is the amount of cash at time $T$ and where $q^\delta_T$ is the remaining quantity of shares at time $T$. In this setting, the trader can sell the shares remaining at time $T$ in his portfolio at a price below the reference price, namely $S_T - \ell(q^\delta_T)$, the function $\ell$ being a positive and increasing penalization function, measuring execution cost.\\

We associate to this stochastic control problem the value function $V_{\Delta}$ defined by:

$$V_\Delta(t,x,q,s) = \sup_{\delta \in \mathcal{A}(t)} \mathbb{E}\left[- \exp\left(-\gamma\left(X^{t,x,\delta}_T+q^{t,q,\delta}_T (S^{t,s}_T-\ell(q^{t,q,\delta}_T))\right)\right) \right],$$
where $\mathcal{A}(t)$ is the set of predictable processes on $[t,T]$, bounded from below and where:
$$dS^{t,s}_\tau = \mu d\tau + \sigma dW_\tau, \qquad S^{t,s}_t = s,$$
$$dX^{t,x,\delta}_\tau = (S_\tau + \delta_\tau)  \Delta dN_\tau, \qquad  X^{t,x,\delta}_t = x,$$
$$dq^{t,q,\delta}_\tau = - \Delta dN_\tau, \qquad  q^{t,q,\delta}_t = q,$$
the point process $N$ having stochastic intensity $(\lambda_\tau)_\tau$ with $\lambda_\tau = \Lambda_\Delta(\delta_\tau) 1_{q_{\tau-} > 0}$.\\

This setting deserves several remarks. First, orders are of constant size $\Delta$. This is a modeling choice corresponding to the way practitioners proceed with liquidation, $\Delta$ being then a fraction of the average trade size (ATS). Also, we implicitly assume that our orders are either entirely filled or not filled at all. In other words, there is no partial fill in this model. This hypothesis is a questionable one since partial fills are common in practice. When using the model in practice, one can always consider a convex combination of optimal quotes between two multiples of $\Delta$. Allowing for partial fills would make the model more realistic. However, it is complicated from a mathematical point of view.\\
A second important point regards negative $\delta$s. We indeed assume that $\Lambda_{\Delta}$ is defined on the entire real line and not only on $\mathbb{R}_+$. Our model allows to post orders at a price below the reference price. If the reference price is the first bid quote, these orders correspond to marketable limit orders. One may then wonder why there is execution uncertainty associated to these orders. An answer is linked to high-frequency traders whose capacity to rapidly cancel trades forces practitioners to use \emph{fill and/or kill} orders or other types of marketable limit orders and not market orders. Also, considering the entire real line allows to introduce indirectly execution costs for liquidity-taking orders. We shall see in Section 6 that there is a link between the execution cost functions of Almgren-Chriss models and the intensity functions on $\lbrace \delta < 0 \rbrace$. It is noteworthy that if one wants to avoid negative $\delta$, adding a hard constraint $\delta \ge 0$ is also possible and does not raise any difficulty. This constrained framework is discussed later in this article (see Section 3.3). To avoid negative $\delta$, some authors (see for instance \cite{bayraktar2011liquidation}) considered an intensity function that blows up at $\delta = 0$. A natural consequence of this modeling choice is that there is unlimited liquidity available at $\delta=0$. This is not a correct approach in our view.\\
The third and last point regards the structural assumption $\Lambda_{\Delta}(\delta) \Lambda''_{\Delta}(\delta) \leq 2{\Lambda'}_{\Delta}(\delta)^2$. This hypothesis, already present in \cite{bayraktar2011liquidation}, is a sufficient condition to guarantee uniqueness of the optimal trading quote. To understand the intuition, let us consider the expected PnL when posting an order at a distance $\delta$ from the reference price. This expected PnL is proportional to $\delta \Lambda_{\Delta}(\delta)$: $\delta$ is the premium over the reference price and $\Lambda_{\Delta}(\delta)$ is the instantaneous probability that a trade takes place at a distance $\delta$ from the reference price. A natural condition for this expression to have a unique maximizer is: $\Lambda_{\Delta} \Lambda''_{\Delta} < 2{\Lambda'}_{\Delta}^2$. In our case, the inequality can be binding because of risk-aversion. It is noteworthy that our framework can be used even if this hypothesis is relaxed.\footnote{The same is true for the assumption $\Lambda_{\Delta} \in C^2$.} However, the consequence is that there may be multiple optimal quotes. For the sake of exposition, we chose to present the model under this structural assumption.

\subsection{A system of ODEs for the value function}

The optimization problem set up in the preceding paragraphs can be solved using classical Bellman tools. To this purpose, we introduce the Hamilton-Jacobi-Bellman equation associated to the optimization problem, where the unknown $u_{\Delta}$ is going to be equal to the value function $V_\Delta$ defined above:

$$(\mathrm{HJB}) \qquad 0 = \partial_t u_\Delta(t,x,q,s) + \mu \partial_{s} u_\Delta(t,x,q,s) +  \frac 12 \sigma^2 \partial_{ss}^2 u_\Delta(t,x,q,s)$$$$ + \sup_{s^a} \Lambda_\Delta(s^a-s) \left[u_\Delta(t,x+\Delta s^a,q-\Delta,s) - u_\Delta(t,x,q,s) \right],$$
with the final condition:
$$u_\Delta(T,x,q,s) = -\exp\left(-\gamma \left(x + q(s-\ell(q)) \right) \right),$$
and the boundary condition:
$$u_\Delta(t,x,0,s) = -\exp\left(-\gamma x \right).$$

Since we use a CARA function, we can factor out the Mark-to-Market (MtM) value $x+qs$ of the portfolio. This remark leads to considering the change of variables $u_{\Delta}(t,x,q,s) = -\exp\left(-\gamma(x+qs+\theta_{\Delta}(t,q))\right)$. In that case, the above HJB equation with 4 variables is (formally) reduced to the following system of ODEs indexed by $q$:

$$ (\mathrm{HJ}_{\theta_\Delta}) \qquad
0=\gamma \partial_t \theta_{\Delta}(t,q) + \gamma \mu q -
\frac{1}{2} \gamma^2 \sigma^2 q^2
+  H_{\Delta}\left( \frac{\theta_{\Delta}(t,q) - \theta_{\Delta}(t,q-\Delta)}{\Delta} \right),
$$
with
$$\theta_{\Delta}(T,q) = -\ell(q) q, \qquad \theta_{\Delta}(t,0) = 0,$$
where
$$H_{\Delta}(p) = \sup_{\delta} \Lambda_{\Delta}(\delta)\left( 1- e^{-\gamma \Delta (\delta - p)}\right).$$

\section{Solution of the optimal control problem}

This section aims at solving the optimal control problem set up in the preceding section. We first concentrate on the equation $(\mathrm{HJ}_{\theta_\Delta})$. Then, we provide a verification theorem that indeed gives a solution to the control problem and characterizes in a simple way the optimal quotes. The last subsection is dedicated to the addition of a hard constraint $\delta \ge \delta^{min}$.

\subsection{A solution to $(\mathrm{HJ}_{\theta_\Delta})$}

We start with a lemma about the hamiltonian function $H_{\Delta}$.

\begin{Lemma}
\label{l1}
Let us define $L_{\Delta}(p,\delta) = \Lambda_{\Delta}(\delta)\left( 1- e^{-\gamma \Delta (\delta - p)}\right)$.

$\forall p \in \mathbb{R}$, $\delta \mapsto L_{\Delta}(p,\delta)$ is strictly increasing on $(-\infty, \tilde{\delta}_{\Delta}^*(p)]$ and strictly decreasing on $[\tilde{\delta}_{\Delta}^*(p),+\infty)$, where $\tilde{\delta}_{\Delta}^*(p)$ is uniquely characterized by $\left(E_{\delta^*_\Delta}\right)$:

$$\tilde{\delta}_{\Delta}^*(p) - \frac{1}{\gamma \Delta} \log\left( 1 - \gamma\Delta \frac{\Lambda_{\Delta} (\tilde{\delta}_{\Delta}^*(p))}{\Lambda'_{\Delta} (\tilde{\delta}_{\Delta}^*(p))}\right) = p, \qquad \left(E_{\delta^*_\Delta}\right).$$

Moreover, $p \mapsto \tilde{\delta}_{\Delta}^*(p)$ is a $C^1$ function.\\

Subsequently, $H_\Delta$ is a $C^1$ function with:

$$H_{\Delta}(p) = \gamma \Delta \frac{\Lambda_{\Delta}(\tilde{\delta}_{\Delta}^*(p))^2}{\gamma\Delta \Lambda_{\Delta}(\tilde{\delta}_{\Delta}^*(p)) - \Lambda'_{\Delta}(\tilde{\delta}_{\Delta}^*(p))}.$$
\end{Lemma}

\textbf{Proof:}\\

Straightforwardly, $\delta \mapsto L_{\Delta}(p,\delta)$ is strictly increasing on $(-\infty, p]$.\\

Regarding the behavior of this function on $[p, +\infty)$, let us notice that $L_{\Delta}(p,p) = 0$ and that $\lim_{\delta \rightarrow +\infty}L_{\Delta}(p,\delta) = 0$.\\

Now, if we differentiate, we get:
$$\partial_\delta L_{\Delta}(p,\delta) = \Lambda'_{\Delta}(\delta)\left( 1- e^{-\gamma \Delta (\delta - p)}\right) +  \gamma \Delta \Lambda_{\Delta}(\delta) e^{-\gamma \Delta (\delta - p)}.$$
Hence, $\partial_\delta L_{\Delta}(p,p) = \gamma \Delta \Lambda_{\Delta}(p) > 0$ and there is at least one $\delta^* \in (p,+\infty)$ such that $\partial_\delta L_{\Delta}(p,\delta^*) = 0$.

Such a $\delta^*$ must satisfy:
$$\delta^* - \frac{1}{\gamma \Delta} \log\left( 1 - \gamma\Delta \frac{\Lambda_{\Delta} (\delta^*)}{\Lambda'_{\Delta} (\delta^*)}\right) = p.$$

Now, $f(x) = x - \frac{1}{\gamma \Delta} \log\left( 1 - \gamma\Delta \frac{\Lambda_{\Delta} (x)}{\Lambda'_{\Delta} (x)}\right)$ defines a strictly increasing function since
\begin{eqnarray*}
  f'(x) &=& 1+\frac{ \left(\frac{\Lambda_{\Delta} (x)}{\Lambda'_{\Delta} (x)}\right)'}{ 1 - \gamma\Delta \frac{\Lambda_{\Delta} (x)}{\Lambda'_{\Delta} (x)}} \\
   &=& 1+\frac{\Lambda'_{\Delta} (x)^2 - \Lambda_{\Delta} (x)\Lambda''_{\Delta} (x) }{ \Lambda'_{\Delta} (x)^2 - \gamma\Delta \Lambda_{\Delta} (x)\Lambda'_{\Delta} (x)} \\
   &=& \frac{- \gamma\Delta \Lambda_{\Delta} (x)\Lambda'_{\Delta} (x)}{\Lambda'_{\Delta} (x)^2 - \gamma\Delta \Lambda_{\Delta} (x)\Lambda'_{\Delta} (x)}+\frac{2\Lambda'_{\Delta} (x)^2 - \Lambda_{\Delta} (x)\Lambda''_{\Delta} (x) }{ \Lambda'_{\Delta} (x)^2 - \gamma\Delta \Lambda_{\Delta} (x)\Lambda'_{\Delta} (x)}\\
\end{eqnarray*}
is strictly positive because of the hypotheses on $\Lambda_\Delta$.\\

Hence $\delta^*$, defined by $f(\delta^*) = p$, is unique and $L_{\Delta}(p,\cdot)$ is strictly increasing on $(-\infty, \tilde{\delta}_{\Delta}^*(p)]$ and strictly decreasing on $[\tilde{\delta}_{\Delta}^*(p),+\infty)$,  where $\tilde{\delta}_{\Delta}^*(p)$ is uniquely characterized by:

$$f(\tilde{\delta}_{\Delta}^*(p)) = \tilde{\delta}_{\Delta}^*(p) - \frac{1}{\gamma \Delta} \log\left( 1 - \gamma\Delta \frac{\Lambda_{\Delta} (\tilde{\delta}_{\Delta}^*(p))}{\Lambda'_{\Delta} (\tilde{\delta}_{\Delta}^*(p))}\right) = p.$$

Using the implicit function theorem, this also gives that $p \mapsto \tilde{\delta}_{\Delta}^*(p)$ is a $C^1$ function.\\

Plugging the relation for $\tilde{\delta}_{\Delta}^*(p)$ in the definition of $H_{\Delta}$ then gives the last part of the lemma.\qed\\

Now, we are going to prove a comparison principle for the system of ODEs. This result is useful in two ways. First, it gives \emph{a priori} bounds that will allow us to prove the existence of a solution to $(\mathrm{HJ}_{\theta_\Delta})$. Second, it will provide bounds to $\theta_\Delta$ independently of $\Delta$ in Section 5, when we shall consider the limiting behavior of $\theta_\Delta$ as $\Delta \to 0$.\\

\begin{Proposition}[Comparison principle]
\label{comp}
Let $\tau \in [0,T)$, $k \in \mathbb{N}$.\\

Let $\underline{\theta}_{\Delta}: [\tau,T]\times \lbrace 0,\Delta,\ldots,k \Delta\rbrace \rightarrow \mathbb{R}$ be a $C^1$ function with respect to time with
$$\forall q\in \lbrace 0,\Delta,\ldots,k \Delta\rbrace,\quad  \underline{\theta}_{\Delta}(T,q) \leq -\ell(q) q \qquad \forall t \in [\tau,T],\quad \underline{\theta}_{\Delta}(t,0) \leq 0,$$
and
$$0 \leq \gamma\partial_t \underline{\theta}_{\Delta}(t,q) + \gamma \mu q -
\frac{1}{2}\gamma^2\sigma^2 q^2
+ H_{\Delta}\left( \frac{\underline{\theta}_{\Delta}(t,q) - \underline{\theta}_{\Delta}(t,q-\Delta)}{\Delta} \right).
$$

Let $\overline{\theta}_{\Delta}: [\tau,T]\times \lbrace 0,\Delta,\ldots,k \Delta\rbrace \rightarrow \mathbb{R}$ be a $C^1$ function with respect to time with
$$\forall q \in \lbrace 0,\Delta,\ldots,k \Delta\rbrace,\quad \overline{\theta}_{\Delta}(T,q) \geq -\ell(q) q \qquad \forall t \in [\tau,T],\quad \overline{\theta}_{\Delta}(t,0) \geq 0,$$
and
$$0 \geq \gamma\partial_t \overline{\theta}_{\Delta}(t,q) + \gamma \mu q -
\frac{1}{2}\gamma^2\sigma^2 q^2
+ H_{\Delta}\left( \frac{\overline{\theta}_{\Delta}(t,q) - \overline{\theta}_{\Delta}(t,q-\Delta)}{\Delta} \right).
$$

Then
$$\overline{\theta}_{\Delta} \geq \underline{\theta}_{\Delta}.$$

\end{Proposition}

\textbf{Proof:}\\

Let $\alpha > 0$.

Let us consider a point $(t^*_{\alpha}, q^*_{\alpha})$ such that
$$ \underline{\theta}_{\Delta}(t^*_{\alpha},q^*_{\alpha}) - \overline{\theta}_{\Delta}(t^*_{\alpha},q^*_{\alpha}) - \alpha(T-t^*_{\alpha}) =
\sup_{(t,q) \in [\tau,T]\times\lbrace0,\ldots,k\Delta\rbrace}\underline{\theta}_{\Delta}(t,q) - \overline{\theta}_{\Delta}(t,q) - \alpha(T-t).
$$

If $t^*_{\alpha}\neq T$ and $q^*_{\alpha}\neq 0$ then:
$$\partial_t \underline{\theta}_{\Delta}(t^*_{\alpha},q^*_{\alpha}) - \partial_t \overline{\theta}_{\Delta}(t^*_{\alpha},q^*_{\alpha}) \leq -\alpha.$$

By definition of $(t^*_{\alpha},q^*_{\alpha})$, since $q^*_{\alpha} \neq 0$:
$$\underline{\theta}_{\Delta}(t^*_{\alpha},q^*_{\alpha}) - \overline{\theta}_{\Delta}(t^*_{\alpha},q^*_{\alpha})  \geq  \underline{\theta}_{\Delta}(t^*_{\alpha},q^*_{\alpha}-\Delta) - \overline{\theta}_{\Delta}(t^*_{\alpha},q^*_{\alpha}-\Delta),$$
and
$$\underline{\theta}_{\Delta}(t^*_{\alpha},q^*_{\alpha}) - \underline{\theta}_{\Delta}(t^*_{\alpha},q^*_{\alpha}-\Delta)   \geq  \overline{\theta}_{\Delta}(t^*_{\alpha},q^*_{\alpha}) - \overline{\theta}_{\Delta}(t^*_{\alpha},q^*_{\alpha}-\Delta).$$

Now, by definition of the functions $\underline{\theta}_{\Delta}$ and $\overline{\theta}_{\Delta}$, we have:
$$ 0 \leq \gamma\left[ \partial_t \underline{\theta}_{\Delta}(t^*_{\alpha},q^*_{\alpha}) - \partial_t \overline{\theta}_{\Delta}(t^*_{\alpha},q^*_{\alpha}) \right] $$
$$+  \left[ H_{\Delta}\left( \frac{\underline{\theta}_{\Delta}(t^*_{\alpha},q^*_{\alpha}) - \underline{\theta}_{\Delta}(t^*_{\alpha},q^*_{\alpha}-\Delta)}{\Delta} \right) - H_{\Delta}\left( \frac{\overline{\theta}_{\Delta}(t^*_{\alpha},q^*_{\alpha}) - \overline{\theta}_{\Delta}(t^*_{\alpha},q^*_{\alpha}-\Delta)}{\Delta} \right)  \right].$$

Since $H_{\Delta}$ is a decreasing function, we have

$$H_{\Delta}\left( \frac{\underline{\theta}_{\Delta}(t^*_{\alpha},q^*_{\alpha}) - \underline{\theta}_{\Delta}(t^*_{\alpha},q^*_{\alpha}-\Delta)}{\Delta} \right) \leq H_{\Delta}\left( \frac{\overline{\theta}_{\Delta}(t^*_{\alpha},q^*_{\alpha}) - \overline{\theta}_{\Delta}(t^*_{\alpha},q^*_{\alpha}-\Delta)}{\Delta} \right).$$

This leads to $0  \leq - \gamma \alpha$ which is not possible.\\

Therefore $t^*_{\alpha} = T$ or $q^*_{\alpha} = 0$,  so that:
$$\sup_{(t,q) \in [\tau,T]\times\lbrace0,\ldots,k\Delta\rbrace}\underline{\theta}_{\Delta}(t,q) - \overline{\theta}_{\Delta}(t,q) - \alpha(T-t)$$$$ = \max\left(\sup_{q \in \lbrace0,\ldots,k\Delta\rbrace}\underline{\theta}_{\Delta}(T,q) - \overline{\theta}_{\Delta}(T,q), \sup_{t\in [\tau,T]}\underline{\theta}_{\Delta}(t,0) - \overline{\theta}_{\Delta}(t,0) - \alpha (T-t) \right)  \leq 0.$$

Thus $\forall (t,q), \underline{\theta}_{\Delta}(t,q) - \overline{\theta}_{\Delta}(t,q) \leq \alpha (T-t) \le \alpha T$. Sending $\alpha$ to $0$ proves our result.\qed\\

We are now ready to prove that the equation $(\mathrm{HJ}_{\theta_\Delta})$ has a unique solution.

\begin{Proposition}
\label{existuniq}
There exists a unique function $\theta_\Delta : [0,T]\times \Delta \mathbb{N} \rightarrow \mathbb{R}$ such that:
\begin{itemize}
  \item $t \in [0,T] \mapsto (\theta_\Delta(\cdot,q))_{q \in \Delta \mathbb{N}}$ is continuously differentiable.
  \item $\theta_\Delta$ is a solution of $(\mathrm{HJ}_{\theta_\Delta})$.
\end{itemize}
\end{Proposition}

\textbf{Proof:}\\

We proceed by induction on $q$. For $q=0$, we have by definition $\theta_{\Delta}(t,0) = 0$.\\

Now, for a given $q \in \Delta \mathbb{N}^*$, let us suppose that $\theta_\Delta(\cdot,q') : [0,T] \to \mathbb{R}$ is a $C^1$ function $\forall q' \le q -\Delta$. Then, the ODE

$$0=\gamma \partial_t \theta_{\Delta}(t,q) + \gamma \mu q - \frac{1}{2} \gamma^2 \sigma^2 q^2 +  H_{\Delta}\left( \frac{\theta_{\Delta}(t,q) - \theta_{\Delta}(t,q-\Delta)}{\Delta} \right),
$$
with the terminal condition $$\theta_{\Delta}(T,q) = -\ell(q) q,$$
satisfies the assumptions of Cauchy-Lipschitz theorem. Consequently, there exists a unique solution $t \mapsto \theta_{\Delta}(t,q)$ on a maximal interval that is a sub-interval of $[0,T]$ and we want to show that this sub-interval is $[0,T]$ itself.\\

To prove this, let suppose by contradiction that $(\tilde{t},T]$ is the maximal interval with $\tilde{t} \ge 0$.\\
Let us notice that, because $H_\Delta$ is positive, $t \mapsto \theta_{\Delta}(t,q) - \mu q(T-t) + \frac{1}{2} \gamma \sigma^2 q^2 (T-t)$ is decreasing. Hence, the only possibility for $(\tilde{t},T]$ to be a maximal interval in $[0,T]$ is that $\lim_{t\to\tilde{t}^+} \theta_{\Delta}(t,q) = +\infty$.\\

Now let us consider $\eta > 0$, $k=\frac q\Delta$ and $\tau = \tilde{t}+\eta$. We define on $[\tau,T]\times\lbrace \Delta, \ldots, q \rbrace$ the two functions $\underline{\theta}_{\Delta}$ and $\overline{\theta}_{\Delta}$ defined by:
$$\underline{\theta}_{\Delta} = \theta_{\Delta}$$
and
$$\forall q' \le q, \overline{\theta}_{\Delta}(t,q') = \mu^+q (T-t) + \frac{1}{\gamma} H_{\Delta}(0) (T-t).$$

These two functions satisfy the assumptions of the above comparison principle. We indeed have that:
$$ \forall q' \le q, \overline{\theta}_{\Delta}(T,q') = 0 \geq -\ell(q') q' \qquad \forall t \in [\tau,T], \overline{\theta}_{\Delta}(t,0) = \mu^+ q (T-t) + \frac{1}{\gamma} H_{\Delta}(0)(T-t) \geq 0,$$
and
$$\forall t \in [\tau,T], \forall q' \in \lbrace \Delta, \ldots, q\rbrace, \gamma\partial_t \overline{\theta}_{\Delta}(t,q') + \gamma \mu q' -
\frac{1}{2}\gamma^2\sigma^2 {q'}^2
+ H_{\Delta}\left( \frac{\overline{\theta}_{\Delta}(t,q') - \overline{\theta}_{\Delta}(t,q'-\Delta)}{\Delta} \right)
$$
$$
= -\gamma \frac{1}{\gamma} H_{\Delta}(0) -\gamma \mu^+ q + \gamma \mu q'  - \frac{1}{2}\gamma^2\sigma^2 {q'}^2 + H_{\Delta}(0)$$$$ = -\gamma (\mu^+ q - \mu q') - \frac{1}{2}\gamma^2\sigma^2 {q'}^2 \le 0.
$$
Hence, $\forall \eta >0, \forall t \in [\tilde{t}+\eta,T], \theta_{\Delta}(t,q) \le \mu^+q (T-\tilde{t}) + \frac{1}{\gamma} H_{\Delta}(0) (T-\tilde{t})$, in contradiction with the fact that $\lim_{t\to\tilde{t}^+} \theta_{\Delta}(t,q) = +\infty$.\\

Hence, $t \mapsto \theta_{\Delta}(t,q)$ is defined on $[0,T]$ and this proves the result.\qed\\

\subsection{Verification theorem and optimal quotes}

Now, we can solve the initial optimal control problem and find the optimal quotes at which the trader should post his limit orders.

\begin{Theorem}[Verification theorem and optimal quotes]
\label{t1}
Let us consider the solution $\theta_\Delta$ of the system $(\mathrm{HJ}_{\theta_\Delta})$.\\
Let us define $u_\Delta(t,x,q,s) = -\exp(-\gamma(x+qs + \theta_\Delta(t,q)))$.\\

We have:
\begin{itemize}
  \item $u_{\Delta}$ is a solution to (HJB),
  \item $u_\Delta$ is equal to the value function $V_{\Delta}$.
\end{itemize}

Moreover, the optimal ask quote $S^a_t = S_t + \delta_{\Delta}^{*}(t)$, for $q_t>0$,  is characterized by:
$$\delta_{\Delta}^{*}(t) = \tilde{\delta}_{\Delta}^*\left(\frac{\theta_\Delta(t,q_t) - \theta_\Delta(t,q_t-\Delta)}{\Delta}\right),$$
where $\tilde{\delta}_{\Delta}^*(\cdot)$ is the function defined in Lemma \ref{l1}.
\end{Theorem}

\textbf{Proof:}\\

From the very definition of $\theta_\Delta$ and $u_\Delta$, it is straightforward to see that $u_\Delta$ is a solution of (HJB).\\

We indeed have that the boundary condition and the terminal condition are satisfied, and for $q \ge \Delta$, we have that:

$$\partial_t u_\Delta(t,x,q,s) + \mu \partial_{s} u_\Delta(t,x,q,s) + \frac 12 \sigma^2 \partial_{ss}^2 u_\Delta(t,x,q,s)$$$$ +\sup_{\delta} \Lambda_\Delta(\delta) \left[u_\Delta(t,x+\Delta s+\Delta\delta,q-\Delta,s) - u_\Delta(t,x,q,s) \right]$$
$$=-\gamma u_\Delta(t,x,q,s) \partial_t \theta_\Delta(t,q) - \gamma \mu q u_\Delta(t,x,q,s)  + \frac 12 \sigma^2 \gamma^2 q^2 u_\Delta(t,x,q,s)$$$$+\sup_{\delta} \Lambda_\Delta(\delta) u_\Delta(t,x,q,s)\left( \exp\left(-\gamma(\Delta \delta -(\theta_\Delta(t,q)- \theta_\Delta(t,q-\Delta)))\right) - 1 \right)$$
$$=-u_\Delta(t,x,q,s)\left[\gamma \partial_t \theta_\Delta(t,q) + \gamma \mu q - \frac 12 \sigma^2 \gamma^2 q^2\right.$$$$\left. + \sup_{\delta} \Lambda_\Delta(\delta) \left( 1- \exp\left(-\gamma(\Delta \delta -(\theta_\Delta(t,q)- \theta_\Delta(t,q-\Delta)))\right) \right) \right]$$
$$=-u_\Delta(t,x,q,s)\left[\gamma \partial_t \theta_\Delta(t,q) + \gamma \mu q - \frac 12 \sigma^2 \gamma^2 q^2 + H_{\Delta}\left(\frac{\theta_\Delta(t,q)- \theta_\Delta(t,q-\Delta)}{\Delta}\right)\right] = 0.$$

Now, we need to verify that $u_\Delta$ is indeed the value function associated to the problem and to prove that our candidate $(\delta_\Delta^{*})_t$ is indeed the optimal control. To that purpose, let us consider a control $\delta \in \mathcal{A}(t)$ and let us consider the following processes for $\tau \in [t,T]$:

$$dS^{t,s}_\tau = \mu d\tau + \sigma dW_\tau, \qquad S^{t,s}_t = s,$$
$$dX^{t,x,\delta}_\tau = (S_\tau + \delta_\tau)  \Delta dN_\tau, \qquad  X^{t,x,\delta}_t = x,$$
$$dq^{t,q,\delta}_\tau = - \Delta dN_\tau, \qquad  q^{t,q,\delta}_t = q,$$
where the point process has stochastic intensity $(\lambda_\tau)_\tau$ with $\lambda_\tau =  \Lambda_\Delta(\delta_\tau) 1_{q_{\tau-} > 0}$.\footnote{This intensity being bounded since $\delta$ is bounded from below.}\\

Now, let us write Itô's formula for $u_\Delta$:\footnote{The equality is still valid when $q_\tau=0$ because of the boundary condition for $u_\Delta$, and because the intensity process is then assumed to be $0$.}

$$u_\Delta(T,X^{t,x,\delta}_{T-},q^{t,q,\delta}_{T-},S^{t,s}_{T}) = u_\Delta(t,x,q,s)$$$$ + \int_t^T \left(\partial_\tau u_\Delta(\tau,X^{t,x,\delta}_{\tau-},q^{t,q,\delta}_{\tau-},S^{t,s}_{\tau})  + \mu \partial_{s} u_\Delta(\tau,X^{t,x,\delta}_{\tau-},q^{t,q,\delta}_{\tau-},S^{t,s}_\tau) + \frac {\sigma^2}2 \partial^2_{ss} u_\Delta(\tau,X^{t,x,\delta}_{\tau-},q^{t,q,\delta}_{\tau-},S^{t,s}_\tau)\right)d\tau$$
$$+ \int_t^T  \left(u_\Delta(\tau,X^{t,x,\delta}_{\tau-}+\Delta S^{t,s}_\tau + \Delta\delta_\tau,q^{t,q,\delta}_{\tau-}-\Delta,S^{t,s}_\tau) - u_\Delta(\tau,X^{t,x,\delta}_{\tau-},q^{t,q,\delta}_{\tau-},S^{t,s}_\tau)\right) \lambda_\tau d\tau
$$$$ +\int_t^T \mu \partial_s u_\Delta(\tau,X^{t,x,\delta}_{\tau-},q^{t,q,\delta}_{\tau-},S^{t,s}_\tau) d{\tau}  +\int_t^T \sigma \partial_s u_\Delta(\tau,X^{t,x,\delta}_{\tau-},q^{t,q,\delta}_{\tau-},S^{t,s}_\tau) dW_{\tau}$$$$ + \int_t^T  \left(u_\Delta(\tau,X^{t,x,\delta}_{\tau-}+ \Delta S^{t,s}_\tau + \Delta\delta_\tau,q^{t,q,\delta}_{\tau-}-\Delta,S^{t,s}_\tau) - u_\Delta(\tau,X^{t,x,\delta}_{\tau-},q^{t,q,\delta}_{\tau-},S^{t,s}_\tau)\right) dM_\tau,$$
where $M$ is the compensated process associated to $N$ for the intensity process $(\lambda_\tau)_\tau$.\\

Now, we have to ensure that the last two integrals consist of martingales so that their mean is $0$. To that purpose, let us notice that $\partial_s u = -\gamma q u$, and hence, since the process $q^{t,q,\delta}$ takes values between $0$ and $q$, we just have to prove that:

$$\mathbb{E}\left[\int_t^T u_\Delta(\tau,X^{t,x,\delta}_{\tau-},q^{t,q,\delta}_{\tau-},S^{t,s}_\tau)^2 d\tau\right] < +\infty,$$

$$\mathbb{E}\left[\int_t^T \left|u_\Delta(\tau,X^{t,x,\delta}_{\tau-}+\Delta S^{t,s}_\tau + \Delta\delta_\tau,q^{t,q,\delta}_{\tau-}-\Delta,S^{t,s}_\tau)\right| \lambda_\tau d\tau\right] < +\infty,$$
and
$$\mathbb{E}\left[\int_t^T \left|u_\Delta(\tau,X^{t,x,\delta}_{\tau-},q^{t,q,\delta}_{\tau-},S^{t,s}_\tau)\right| \lambda_\tau d\tau\right] < +\infty.$$

We have:

$$u_\Delta(\tau,X^{t,x,\delta}_\tau,q^{t,q,\delta}_\tau,S^{t,s}_\tau)^2 \le \exp\left(2\gamma \|\theta_\Delta\|_{\infty}\right)\exp{\left(-2\gamma(X^{t,x,\delta}_\tau+q^{t,q,\delta}_\tau S^{t,s}_\tau)\right)}$$
$$\le  \exp\left(2\gamma \|\theta_\Delta\|_{\infty}\right)\exp{\left(-2\gamma(x-q\|\delta^-\|_{\infty} + 2q \inf_{\tau \in [t,T]} S^{t,s}_\tau 1_{\inf_{\tau \in [t,T]} S^{t,s}_\tau < 0} )\right)}$$
$$\le \exp\left(2\gamma \|\theta_\Delta\|_{\infty}\right)\exp{\left({-2\gamma(x-q\|\delta^-\|_{\infty})}\right)} \left(1+\exp{\left(-2\gamma q\inf_{\tau \in [t,T]} S^{t,s}_\tau\right)} \right).$$
Hence:
$$\mathbb{E}\left[\int_t^T u_\Delta(\tau,X^{t,x,\delta}_\tau,q^{t,q,\delta}_\tau,S^{t,s}_\tau)^2 d\tau\right] = \mathbb{E}\left[\int_t^T u_\Delta(\tau,X^{t,x,\delta}_{\tau-},q^{t,q,\delta}_{\tau-},S^{t,s}_\tau)^2 d\tau\right]$$
$$\le \exp\left(2\gamma \|\theta_\Delta\|_{\infty}\right) \exp{\left({-2\gamma(x-q\|\delta^-\|_{\infty})}\right)} (T-t) \left(1+\mathbb{E}\left[\exp{\left(-2\gamma q\inf_{\tau \in [t,T]} S^{t,s}_\tau\right)} \right]\right) < +\infty,$$
because of the law of $\inf_{\tau \in [t,T]} S^{t,s}_\tau$.\\

Now, the same argument works for the second and third integrals, noticing that $\delta$ is bounded from below and that $\lambda$ is bounded.\\

Hence, since we have, by construction\footnote{This inequality is also true when the portfolio is empty because of the boundary conditions.}

$$\partial_\tau u_\Delta(\tau,X^{t,x,\delta}_{\tau-},q^{t,q,\delta}_{\tau-},S^{t,s}_\tau) + \mu \partial_{s} u_\Delta(\tau,X^{t,x,\delta}_{\tau-},q^{t,q,\delta}_{\tau-},S^{t,s}_\tau) + \frac {\sigma^2}2 \partial^2_{ss} u_\Delta(\tau,X^{t,x,\delta}_{\tau-},q^{t,q,\delta}_{\tau-},S^{t,s}_\tau)$$$$+ \left(u_\Delta(\tau,X^{t,x,\delta}_{\tau-}+\Delta S^{t,s}_\tau + \Delta \delta_t,q^{t,q,\delta}_{\tau-}-\Delta,S^{t,s}_\tau) - u_\Delta(\tau,X^{t,x,\delta}_{\tau-},q^{t,q,\delta}_{\tau-},S^{t,s}_\tau)\right) \lambda_\tau \le 0,$$

we obtain that

$$\mathbb{E}\left[u_\Delta(T,X^{t,x,\delta}_T,q^{t,q,\delta}_T,S^{t,s}_T)\right] = \mathbb{E}\left[u_\Delta(T,X^{t,x,\delta}_{T-},q^{t,q,\delta}_{T-},S^{t,s}_T)\right] \le u_\Delta(t,x,q,s),$$
and this is true for all $\delta \in \mathcal{A}(t)$. Since for $\delta_t = \delta_{\Delta}^{*}(t)$ we have an equality in the above inequality by construction of the function $\tilde{\delta}_{\Delta}^*$, we obtain that:

$$\sup_{\delta \in \mathcal{A}(t)} \mathbb{E}\left[u_\Delta(T,X^{t,x,\delta}_T,q^{t,q,\delta}_T,S^{t,s}_T)\right] \le u_\Delta(t,x,q,s) = \mathbb{E}\left[u_\Delta(T,X^{t,x,\delta_{\Delta}^{*}}_T,q^{t,q,\delta_{\Delta}^{*}}_T,S^{t,s}_T)\right],$$
\emph{i.e.}
$$\sup_{\delta \in \mathcal{A}(t)} \mathbb{E}\left[-\exp\left(-\gamma\left(X^{t,x,\delta}_T+q^{t,q,\delta}_T(S^{t,s}_T - \ell(q^{t,q,\delta}_T))\right)\right)\right]$$$$ \le u_\Delta(t,x,q,s) = \mathbb{E}\left[-\exp\left(-\gamma\left(X^{t,x,\delta_{\Delta}^{*}}_T+q^{t,q,\delta_{\Delta}^{*}}_T(S^{t,s}_T - \ell(q^{t,q,\delta_{\Delta}^{*}}_T))\right)\right)\right].$$

This proves that $u_\Delta$ is the value function and that $t \mapsto \delta_{\Delta}^{*}(t)$ is optimal.\qed\\

Theorem \ref{t1} proves that the optimal quotes are deterministic. This is linked to the use of a CARA utility function as in the usual Almgren-Chriss framework. Theorem \ref{t1} also provides a simple way to compute the optimal quotes. One has indeed to solve the triangular system of ODEs $(\mathrm{HJ}_{\theta_\Delta})$ to obtain the function $\theta_\Delta$. Numerically, this does not constitute any difficulty and one may use for instance a Euler scheme. Then, once $\theta_\Delta$ has been computed, the optimal quotes are given by the simple expression $\tilde{\delta}_{\Delta}^*\left(\frac{\theta_\Delta(t,q_t) - \theta_\Delta(t,q_t-\Delta)}{\Delta}\right)$ where the function $\tilde{\delta}_{\Delta}^*$ is implicitly characterized by the equation $\left(E_{\delta^*_\Delta}\right)$ of Lemma \ref{l1}, and can be easily computed using Newton's method for instance.\\

\subsection{Introducing a hard constraint $\delta \ge \delta^{min}$}

In the above framework, $\delta$ was allowed to take any value on the real line. To avoid marketable limit orders, one might want to impose a constraint $\delta \ge \delta^{min}$ where $\delta^{min}$ would be positive. Using the same tools as above, the problem with the additional constraint $\delta \ge \delta^{min}$ can be solved easily.\\

The (HJB) equation becomes $$ 0 = \partial_t u^{min}_\Delta(t,x,q,s) + \mu \partial_{s} u^{min}_\Delta(t,x,q,s) +  \frac 12 \sigma^2 \partial_{ss}^2 u^{min}_\Delta(t,x,q,s)$$$$ + \sup_{s^a \ge s + \delta^{min}} \Lambda_\Delta(s^a-s) \left[u^{min}_\Delta(t,x+\Delta s^a,q-\Delta,s) - u^{min}_\Delta(t,x,q,s) \right],$$
with the final condition:
$$u^{min}_\Delta(T,x,q,s) = -\exp\left(-\gamma \left(x + q(s-\ell(q)) \right) \right),$$
and the boundary condition:
$$u^{min}_\Delta(t,x,0,s) = -\exp\left(-\gamma x \right).$$

We consider the change of variables $u^{min}_{\Delta}(t,x,q,s) = -\exp\left(-\gamma(x+qs+\theta^{min}_{\Delta}(t,q))\right)$, as above. We then obtain the following system of ODEs indexed by $q$:

$$ (\mathrm{HJ}_{\theta^{min}_\Delta}) \qquad
0=\gamma \partial_t \theta^{min}_{\Delta}(t,q) + \gamma \mu q -
\frac{1}{2} \gamma^2 \sigma^2 q^2
+  H^{min}_{\Delta}\left( \frac{\theta^{min}_{\Delta}(t,q) - \theta^{min}_{\Delta}(t,q-\Delta)}{\Delta} \right),
$$
with
$$\theta^{min}_{\Delta}(T,q) = -\ell(q) q, \qquad \theta^{min}_{\Delta}(t,0) = 0,$$
where
$$H^{min}_{\Delta}(p) = \sup_{\delta\ge \delta^{min}} \Lambda_{\Delta}(\delta)\left( 1- e^{-\gamma \Delta (\delta - p)}\right) = \sup_{\delta\ge \delta^{min}} L_\Delta(p,\delta).$$

The important point here is to recall that $\delta \mapsto L_\Delta(p,\delta)$ is strictly increasing on $(-\infty, \tilde{\delta}_{\Delta}^*(p)]$ and strictly decreasing on $[\tilde{\delta}_{\Delta}^*(p),+\infty)$. Hence, the unique maximizer of $\delta \mapsto L_\Delta(p,\delta)$ over $\lbrace \delta \ge \delta^{min} \rbrace$ is $\max(\delta^{min}, \tilde{\delta}_{\Delta}^*(p))$.\\

Let us define $p^{min} = \delta^{min} - \frac{1}{\gamma \Delta} \log\left( 1 - \gamma\Delta \frac{\Lambda_{\Delta} (\delta^{min})}{\Lambda'_{\Delta} (\delta^{min})}\right)$. The hamiltonian function $H^{min}_{\Delta}$ can be written:

$$H^{min}_{\Delta}(p) = \begin{cases} H_\Delta(p) &\mbox{if } p \ge p^{min} \\
\Lambda_{\Delta}(\delta^{min})\left( 1- e^{-\gamma \Delta (\delta^{min} - p)}\right) & \mbox{if } p \le p^{min}. \end{cases} $$

It is a locally Lipschitz and decreasing function. In particular, the counterpart of Proposition \ref{comp} and Proposition \ref{existuniq} holds: there exists a unique $C^1$ function $\theta^{min}_{\Delta}$ solution of $(\mathrm{HJ}_{\theta^{min}_\Delta})$. Therefore, we can enounce a verification theorem and find the optimal quotes. The proof is \emph{mutatis mutandis} the same as for Theorem \ref{t1}.

\begin{Theorem}[Verification theorem and optimal quotes]

Let us consider the solution $\theta^{min}_\Delta$ of the system $(\mathrm{HJ}_{\theta^{min}_\Delta})$.\\
Then:
$$-\exp(-\gamma(x+qs + \theta^{min}_\Delta(t,q))) = \sup_{\delta \in \mathcal{A}^{min}(t)} \mathbb{E}\left[- \exp\left(-\gamma\left(X^{t,x,\delta}_T+q^{t,q,\delta}_T (S^{t,s}_T-\ell(q^{t,q,\delta}_T))\right)\right) \right],$$
where $\mathcal{A}^{min}(t)$ is the set of predictable processes on $[t,T]$, bounded from below by $\delta^{min}$ and where:
$$dS^{t,s}_\tau = \mu d\tau + \sigma dW_\tau, \qquad S^{t,s}_t = s,$$
$$dX^{t,x,\delta}_\tau = (S_\tau + \delta_\tau)  \Delta dN_\tau, \qquad  X^{t,x,\delta}_t = x,$$
$$dq^{t,q,\delta}_\tau = - \Delta dN_\tau, \qquad  q^{t,q,\delta}_t = q,$$
the point process $N$ having stochastic intensity $(\lambda_\tau)_\tau$ with $\lambda_\tau = \Lambda_\Delta(\delta_\tau) 1_{q_{\tau-} > 0}$.\\

Moreover, the optimal ask quote $S^a_t = S_t + \delta^{min*}_{\Delta}(t)$, for $q_t>0$,  is characterized by:
$$\delta^{min*}_{\Delta}(t) = \max\left(\delta^{min},\tilde{\delta}_{\Delta}^*\left(\frac{\theta^{min}_\Delta(t,q_t) - \theta^{min}_\Delta(t,q_t-\Delta)}{\Delta}\right)\right),$$ where $\tilde{\delta}_{\Delta}^*(\cdot)$ is the function defined in Lemma \ref{l1}.
\end{Theorem}

\section{Examples and properties}

\subsection{The case of an exponential intensity function}

In the above section, we generalized a model already used in \cite{GLFT}, in which the intensity functions had exponential shape: $\Lambda_{\Delta}(\delta) = A_{\Delta} e^{-k_{\Delta}\delta}$.\\

In the case of exponential intensity, we can write the results of \cite{GLFT} (in a slightly more general case than in the original paper) in the language of this paper. In fact, the reason why closed-form solutions can be obtained in the exponential case is that the equation ($\mathrm{HJ}_{\theta_\Delta}$) simplifies to a linear system of equations when we replace the unknown $\theta_\Delta$ by $\exp\left({\frac{k_\Delta}{\Delta} \theta_\Delta}\right)$:

\begin{Proposition}
Assume that $\Lambda_{\Delta}(\delta) = A_{\Delta} e^{-k_{\Delta}\delta}$.\\

Then, $H_{\Delta}(p) = \frac{\gamma \Delta}{k_{\Delta}}\left(1+\frac{\gamma\Delta}{k_\Delta}\right)^{-1-\frac{k_\Delta}{\gamma\Delta}} A_{\Delta}  e^{-k_{\Delta}p}$.\\

Also, if we consider  $\theta_\Delta : [0,T]\times \Delta \mathbb{N} \rightarrow \mathbb{R}$, the unique $C^1$ solution of $(\mathrm{HJ}_{\theta_\Delta})$, then $w_\Delta = \exp\left({\frac{k_\Delta}{\Delta} \theta_\Delta}\right)$ is the unique solution of:
$$
\partial_t w_{\Delta}(t,q) = - \frac{1}{\Delta} k_\Delta \mu q w_{\Delta}(t,q)+
\frac{1}{2\Delta} \gamma k_\Delta \sigma^2 q^2 w_{\Delta}(t,q)
- A_\Delta \left(1+\frac{\gamma\Delta}{k_\Delta}\right)^{-1-\frac{k_\Delta}{\gamma\Delta}} w_{\Delta}(t,q-\Delta),
$$
with
$$w_{\Delta}(T,q) = e^{- \frac{k_\Delta}{\Delta} \ell(q) q}, \qquad w_{\Delta}(t,0) = 1,$$

and the optimal quote, for $q_t>0$, is given by:

$$\delta_{\Delta}^{*}(t) = \frac 1{k_\Delta}\log\left(\frac{w_\Delta(t,q_t)}{w_\Delta(t,q_t-\Delta)}\right) + \frac 1{\gamma\Delta} \log\left(1+\frac{\gamma\Delta}{k_\Delta}\right).$$

\end{Proposition}

\subsection{Numerical examples}

\begin{figure}[!h]
\center
  \includegraphics[width=420pt]{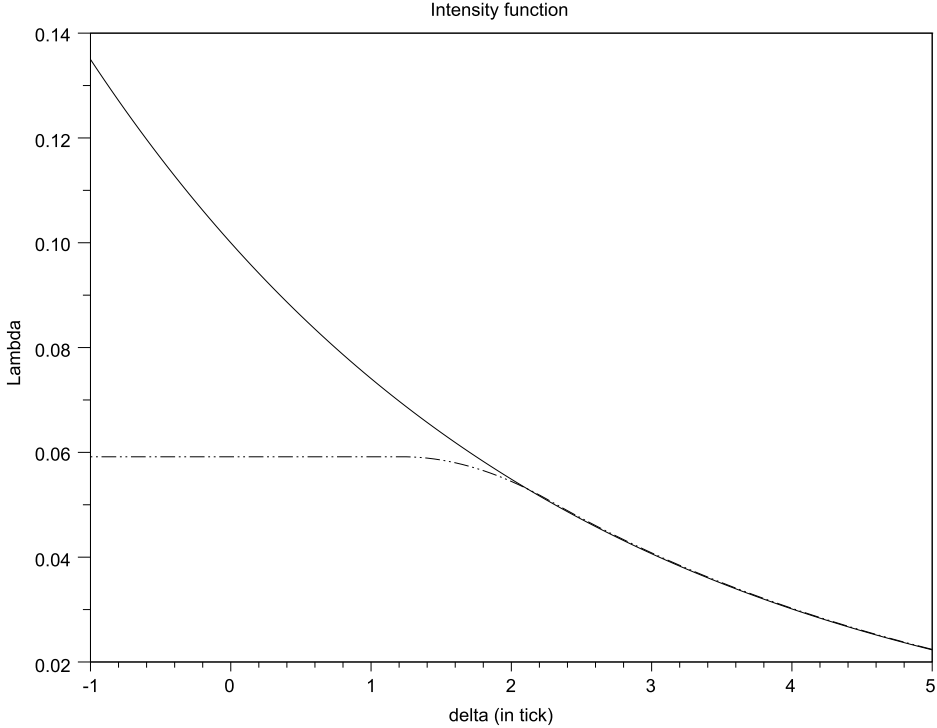}\\
  \caption{Intensity functions (here represented on $[-1,5]$). Line: $\Lambda_{\Delta}(\delta) = A e^{-k\delta}$, $A = 0.1 \mathrm{\;} (\mathrm{s}^{-1})$, $k = 0.3 \mathrm{\;} (\mathrm{Tick}^{-1})$. Dotted line: $\tilde{\Lambda}_{\Delta}(\delta)$, identical to $\Lambda_{\Delta}(\delta)$ for $\delta \ge 2$.}
  \label{lambda}
\end{figure}

We now provide numerical approximations of both the function $\theta_\Delta(t,q)$ and the optimal control function $\delta_\Delta^*(t,q)$. These numerical approximations allow to compare what happens in the pure exponential case and what happens when another intensity function is considered, especially for negative $\delta$s. In this section, we consider as an alternative to the exponential form for $\Lambda_{\Delta}$ a functional form $\tilde{\Lambda}_{\Delta}$ (see Figure \ref{lambda}) that prevents the use of marketable limit orders. The intensity function $\tilde{\Lambda}_{\Delta}$ prevents the use of marketable limit orders since $\tilde{\Lambda}_{\Delta}$ is constant (in fact, decreasing very slowly to satisfy the hypotheses of the paper) for negative $\delta$. Also, we included the commonly observed fact that the probability to be executed does not correspond to the exponential intensity framework for small positive $\delta$s.\\

\begin{figure}[!htbp]
\center
  \includegraphics[width=420pt]{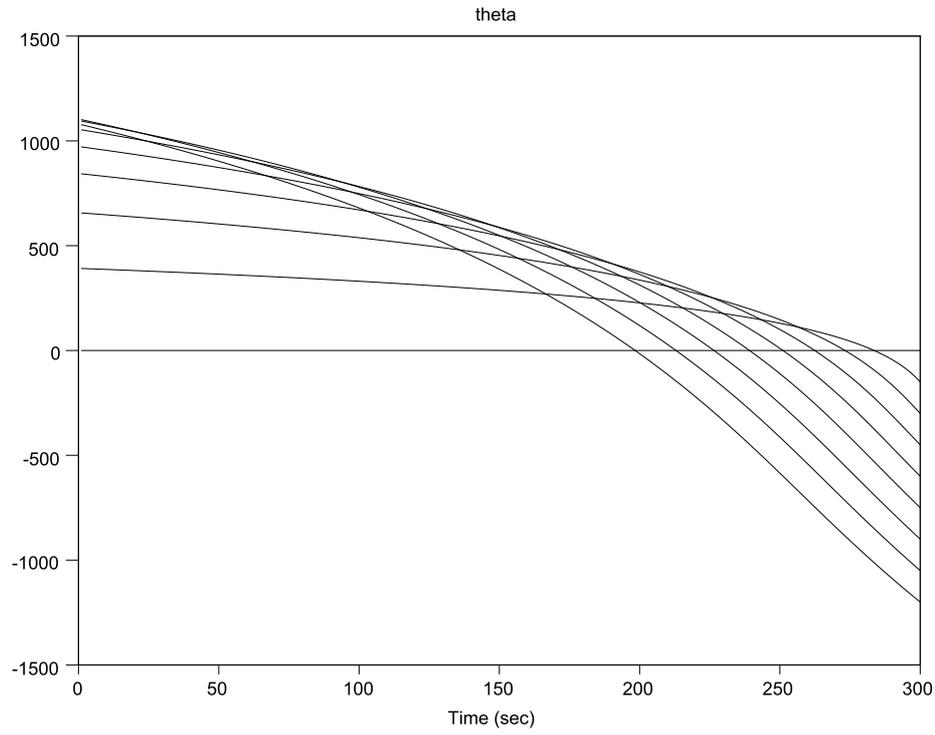}\\
  \includegraphics[width=420pt]{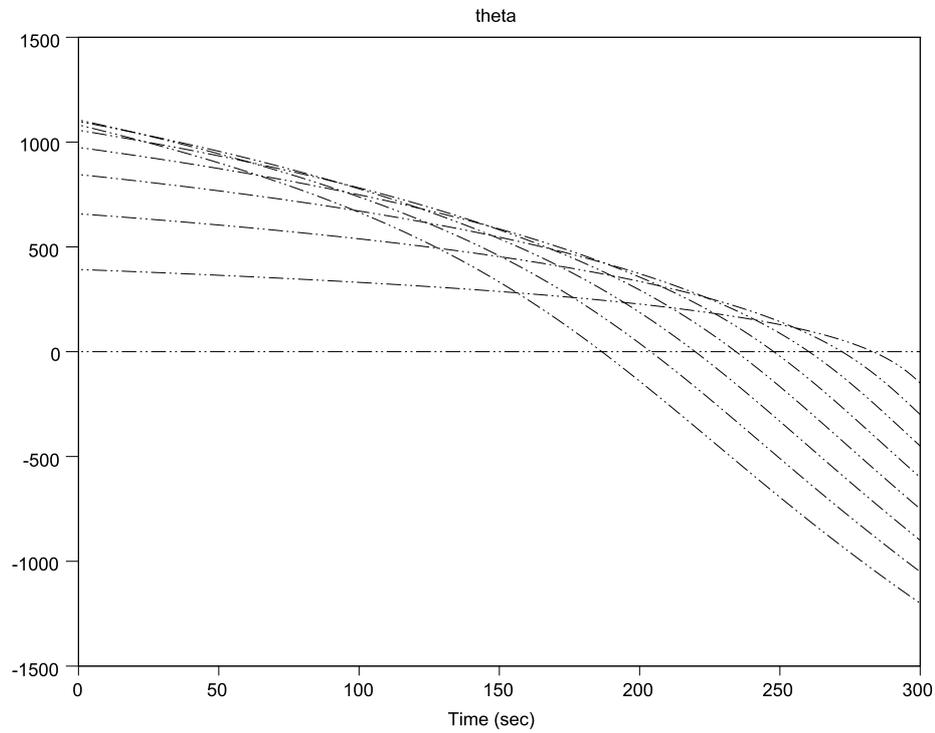}\\
  \caption{Solution $\theta_\Delta(t,q)$ for $\Lambda_{\Delta}$ (top) and $\tilde{\Lambda}_{\Delta}$ (down), $q_0=400$, $\Delta = 50$, $T=300 \mathrm{\;} (\mathrm{s})$, $\mu = 0$, $\sigma = 0.3 \mathrm{\;} (\mathrm{ Tick}.\mathrm{s}^{-\frac{1}{2}})$, $\gamma =  0.001\mathrm{\;}(\mathrm{Tick}^{-1})$ and $\ell(q) = \ell =3\mathrm{\;}(\mathrm{Tick})$. The index $q \in \lbrace 0, 50, \ldots, 400 \rbrace$ of each curve can be read from the terminal values. }
  \label{theta}
\end{figure}

\begin{figure}[!h]
\center
  \includegraphics[width=425pt]{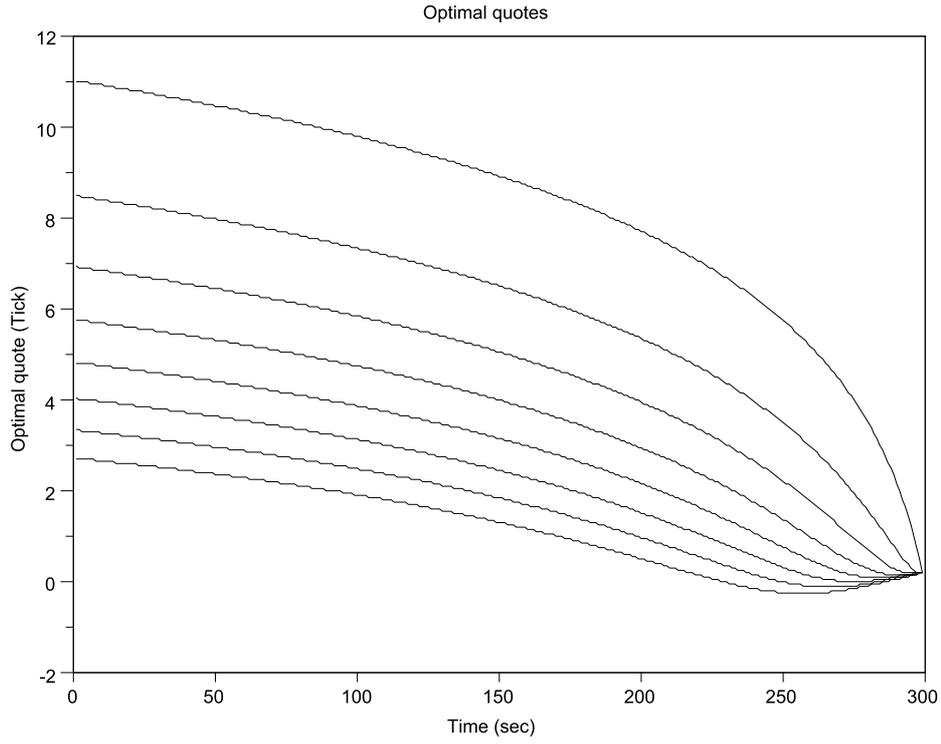}\\
  \includegraphics[width=425pt]{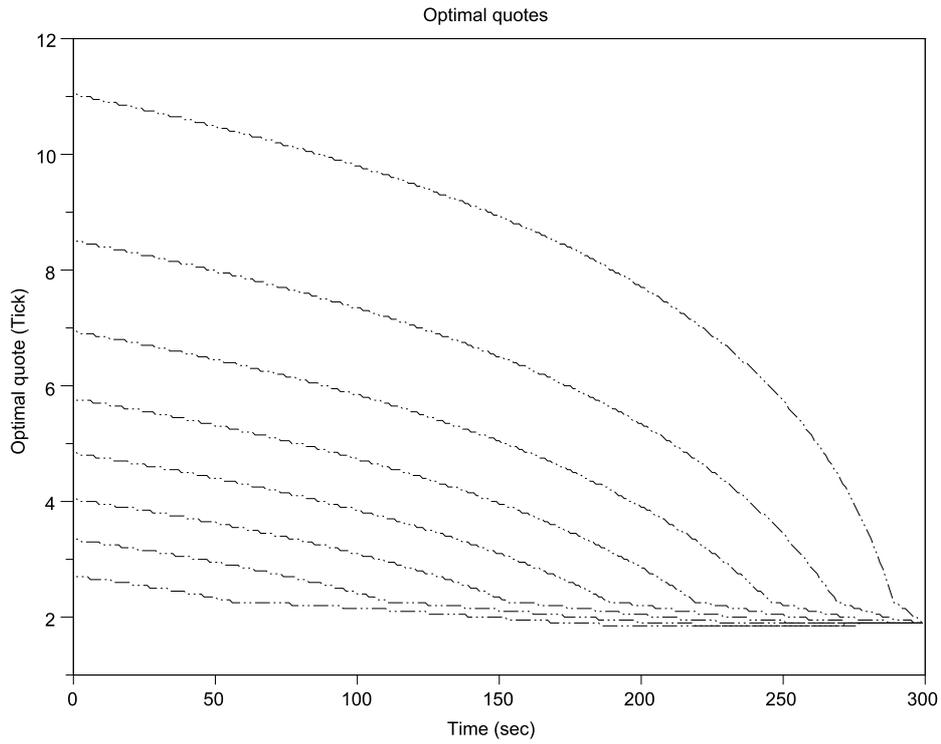}\\
  \caption{Solution $\delta^*_\Delta(t,q)$ for $\Lambda_{\Delta}$ (top) and $\tilde{\Lambda}_{\Delta}$ (down), $q_0=400$, $\Delta = 50$, $T=300 \mathrm{\;} (\mathrm{s})$, $\mu = 0$, $\sigma = 0.3 \mathrm{\;} (\mathrm{ Tick}.\mathrm{s}^{-\frac{1}{2}})$, $\gamma =  0.001\mathrm{\;}(\mathrm{Tick}^{-1})$ and $\ell(q) = \ell =3\mathrm{\;}(\mathrm{Tick})$. The lower the quotes, the higher $q \in \lbrace 50, \ldots, 400 \rbrace$. }
  \label{delta}
\end{figure}

Figure \ref{theta} and Figure \ref{delta} represent respectively the solution $\theta_\Delta$ and the optimal quotes $\delta^*_\Delta(t,q)$ as given by Theorem \ref{t1}.\footnote{One may wonder why we choose a risk aversion parameter $\gamma =  0.001$. This figure seems small but it has in fact an important impact since the shares are sold by groups of $50$.} From Figure \ref{theta}, we know that $\theta_\Delta$ is not a monotonic function of $q$. It is important here to recall the economic meaning of $\theta_\Delta$. The certainty equivalent of holding $q$ shares at time $t$ is $qs+\theta_{\Delta}(t,q)$. Hence, $\theta_\Delta(t,q)$ is a risk-adjusted value of holding $q$ shares at time $t$ in excess of the MtM value $qs$. The reason why $\theta_\Delta(t,q)$ is not a monotonic function of $q$ can then be understood easily. At the time horizon $T$, the function is decreasing but far from $T$ two effects are at stake. On the one hand, when there are many shares in the portfolio, there will be many trades and hence more opportunities to make money through limit orders: this goes in the direction of an increasing function $\theta_\Delta(t,\cdot)$. On the other hand, the larger the inventory to liquidate, the more price risk. This goes in the direction of a decreasing function $\theta_\Delta(t,\cdot)$ since it is a risk-adjusted value.\\

Although, there is almost no difference between the two cases $\Lambda_{\Delta}$ and $\tilde{\Lambda}_{\Delta}$ as far as $\theta_\Delta$ is concerned, this is not true anymore when it comes to the optimal quotes $\delta^*_\Delta$. We indeed see on Figure \ref{delta}, as expected, that, in the case of the intensity function $\tilde{\Lambda}_{\Delta}$, there is no negative optimal quotes. This very conservative choice for the intensity function is a way to avoid any influence of the intensity function on the set $\lbrace\delta < 0\rbrace$. Also, we see that, since $\tilde{\Lambda}_{\Delta}$ is not of exponential form for small positive $\delta$s, the lower bound for the optimal quotes is higher than expected.\\

\begin{figure}[!h]
\center
  \includegraphics[width=420pt]{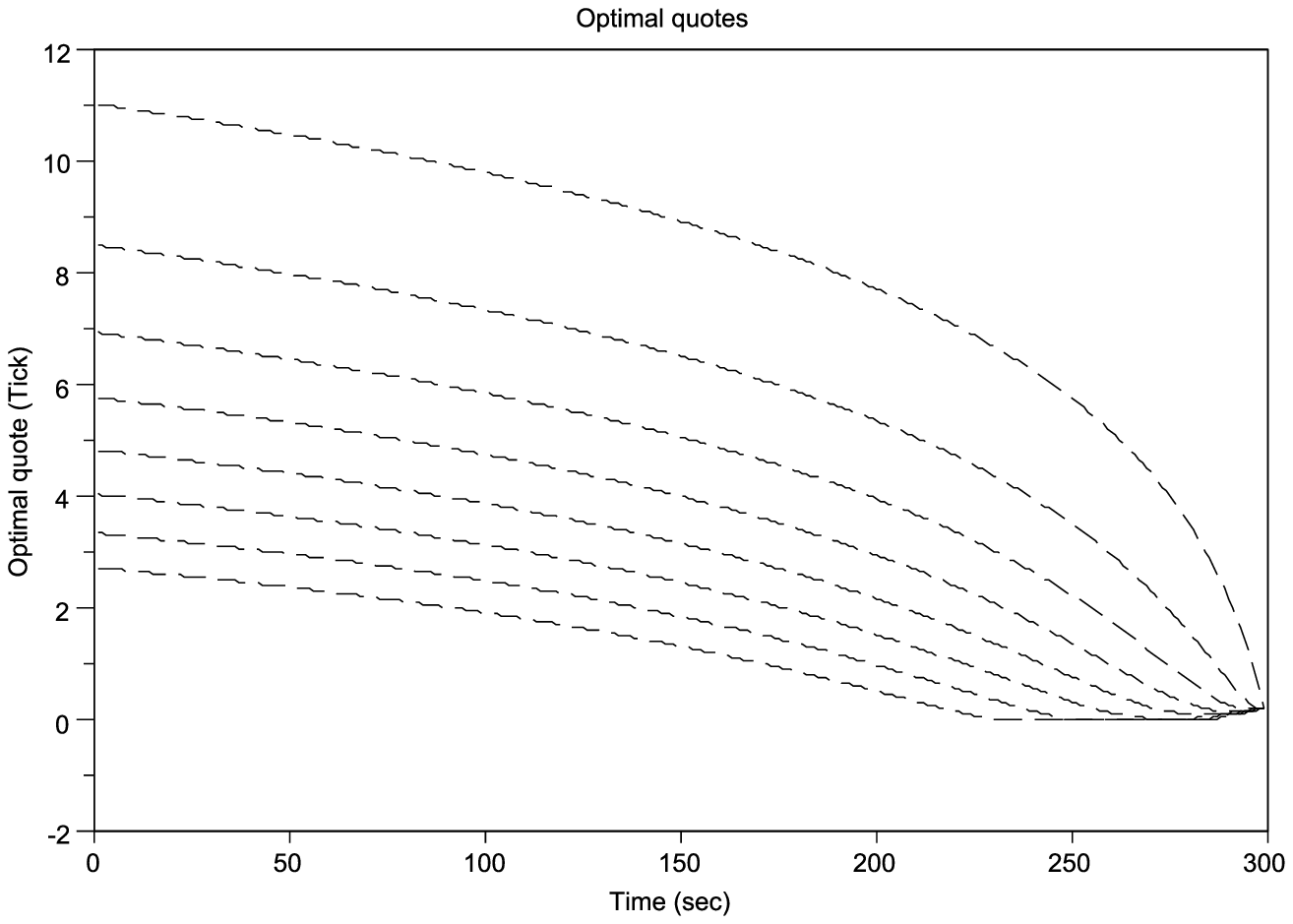}\\
  \caption{Optimal quote for $\Lambda_{\Delta}$ when the constraint $\delta \ge 0$ is imposed. $q_0=400$, $\Delta = 50$, $T=300 \mathrm{\;} (\mathrm{s})$, $\mu = 0$, $\sigma = 0.3 \mathrm{\;} (\mathrm{ Tick}.\mathrm{s}^{-\frac{1}{2}})$, $\gamma =  0.001\mathrm{\;}(\mathrm{Tick}^{-1})$ and $\ell(q) = \ell =3\mathrm{\;}(\mathrm{Tick})$. The lower the quotes, the higher $q \in \lbrace 50, \ldots, 400 \rbrace$. }
  \label{deltapos}
\end{figure}

Another way to prevent marketable limit orders is to impose $\delta \ge \delta^{min}=0$ as in Section 3.3. In that case, if we consider the intensity function $\Lambda_\Delta$ and the same parameters as above, we obtain the quotes given on Figure \ref{deltapos}. These quotes are almost exactly the same as if we had floored the optimal quote $\delta^*_\Delta$ of the unconstrained problem to 0.

\subsection{Asymptotic quote}

In \cite{GLFT}, we obtained a limiting regime when $T \to \infty$. This result generalizes to our general framework. More exactly, we obtain:

\begin{Proposition}[Asymptotic behavior]
Let us suppose that:
\begin{itemize}
\item $\lim_{p\to +\infty} H_{\Delta}(p) = 0$. \footnote{This is guaranteed if $\lim_{\delta \to +\infty} \delta \Lambda_{\Delta}(\delta) = 0$.}
\item $\gamma > 0$,
\item $\mu < \frac 12 \gamma \sigma^2 \Delta$.\\
\end{itemize}
Then, the asymptotic behavior of $\theta_{\Delta}$ is:
$$\lim_{T\to +\infty} \theta_{\Delta}(0,q) = \Delta \sum_{q' \in \lbrace \Delta, 2\Delta, \ldots, q \rbrace} H_{\Delta}^{-1}\left(\frac 12 \gamma^2 \sigma^2 {q'}^2 - \gamma \mu q'\right) = \theta^\infty_{\Delta}(q).$$
The resulting asymptotic behavior of the optimal quote is:
$$\lim_{T\to +\infty} {\delta_\Delta^{*}}(t=0) = \delta_\Delta^{*}\left(H_{\Delta}^{-1}\left(\frac 12 \gamma^2 \sigma^2 q_0^2 - \gamma \mu q_0\right)\right) = \delta_\Delta^{*\infty}.$$
\end{Proposition}

\textbf{Proof:}\\

Let us define for $q\in \Delta \mathbb{N}$:
$$\theta^\infty_{\Delta}(q) = \Delta \sum_{q' \in \lbrace \Delta, 2\Delta, \ldots, q \rbrace} H_{\Delta}^{-1}\left(\frac 12 \gamma^2 \sigma^2 {q'}^2 -\gamma \mu q'\right).\footnote{This is well-defined because of the assumptions on $H_{\Delta}$ and on the parameters, and because $\gamma >0$. In the risk-neutral case, there is no upper bound for the optimal quotes when $T \to \infty$. This constitutes an important difference between our model and the model of Bayraktar and Ludkovski \cite{bayraktar2011liquidation}.}$$

Let us define for $t \ge 0$ and $q \in \Delta \mathbb{N}$, $\theta_{\Delta}^r$, the unique solution\footnote{To prove that this function is well-defined, one can use the same tools as in Proposition \ref{existuniq}.} of:

$$ 0= -\gamma \partial^r_t \theta_{\Delta}(t,q) + \gamma \mu q -
\frac{1}{2} \gamma^2 \sigma^2 q^2
+  H_{\Delta}\left( \frac{\theta_{\Delta}^r(t,q) - \theta^r_{\Delta}(t,q-\Delta)}{\Delta} \right),
$$
with
$$\theta^r_{\Delta}(0,q) = -\ell(q) q, \qquad \theta^r_{\Delta}(t,0) = 0.$$

Then, because we just reversed time, we want to prove that:
$$\forall q \in \Delta \mathbb{N}, \quad \lim_{t \to +\infty} \theta_{\Delta}^r(t,q) = \theta^\infty_{\Delta}(q).$$

We proceed by induction. The result is true for $q=0$. Let us suppose that the result is true for $q-\Delta$ for some $q \in \Delta \mathbb{N}^*$.\\
Then: $$\forall \epsilon > 0, \exists t_{q-\Delta}, \forall t\ge t_{q-\Delta}, |\theta_{\Delta}^r(t,q-\Delta) - \theta^\infty_{\Delta}(q-\Delta)| \le \epsilon.$$

Since $H_\Delta$ is a strictly decreasing function, we obtain that $\forall t\ge t_{q-\Delta}$:

$$\gamma \mu q -\frac 12 \gamma^2 \sigma^2 q^2 + H_{\Delta}\left(\frac{\theta^r_\Delta(t,q) - \theta^\infty_{\Delta}(q-\Delta) + \epsilon }{\Delta}\right) $$$$\le \gamma \partial_t \theta^r_\Delta(t,q) \le \gamma \mu q -\frac 12 \gamma^2 \sigma^2 q^2 + H_{\Delta}\left(\frac{\theta^r_\Delta(t,q) - \theta^\infty_{\Delta}(q-\Delta) - \epsilon }{\Delta}\right),$$
or equivalently:
$$ H_{\Delta}\left(\frac{\theta^r_\Delta(t,q) - \theta^\infty_{\Delta}(q-\Delta) + \epsilon }{\Delta}\right) - H_{\Delta}\left(\frac{\theta^\infty_\Delta(q) - \theta^\infty_{\Delta}(q-\Delta)}{\Delta}\right) $$$$\le \gamma \partial_t \theta^r_\Delta(t,q) \le H_{\Delta}\left(\frac{\theta^r_\Delta(t,q) - \theta^\infty_{\Delta}(q-\Delta) - \epsilon }{\Delta}\right) - H_{\Delta}\left(\frac{\theta^\infty_\Delta(q) - \theta^\infty_{\Delta}(q-\Delta)}{\Delta}\right).$$

Hence, $\forall t\ge t_{q-\Delta}$:

$$ \theta^r_\Delta(t,q) > \theta^\infty_{\Delta}(q) + \epsilon \Rightarrow \partial_t \theta^r_\Delta(t,q) < 0,$$
and
$$ \theta^r_\Delta(t,q) < \theta^\infty_{\Delta}(q) - \epsilon \Rightarrow \partial_t \theta^r_\Delta(t,q) > 0.$$

As a consequence, if there exists $t' \ge t_{q-\Delta}$ such that $|\theta^r_\Delta(t',q) - \theta^\infty_{\Delta}(q)| \le \epsilon$ then, $\forall t \ge t', |\theta^r_\Delta(t,q) - \theta^\infty_{\Delta}(q)| \le \epsilon $.\\

In particular, if $|\theta^r_\Delta(t_{q-\Delta},q) - \theta^\infty_{\Delta}(q)| \le \epsilon$ then, $\forall t \ge t_{q-\Delta}, |\theta^r_\Delta(t,q) - \theta^\infty_{\Delta}(q)| \le \epsilon $.\\

Now, if $\theta^r_\Delta(t_{q-\Delta},q) > \theta^\infty_{\Delta}(q) + \epsilon$, then there are two possibilities. The first one is that the function $t \ge t_{q-\Delta} \mapsto \theta^r_\Delta(t,q)$ is decreasing and in that case it is bounded from below by $\theta^\infty_{\Delta}(q) - \epsilon$ and must converge. Since $\lim_{t \to +\infty} \theta_{\Delta}^r(t,q-\Delta) = \theta^\infty_{\Delta}(q-\Delta)$, the only possible limit for $\theta_{\Delta}^r(t,q)$ is $\theta^\infty_{\Delta}(q)$. The second possibility is that $t \ge t_{q-\Delta} \mapsto \theta^r_\Delta(t,q)$ is not a decreasing function and in that case there must exists $t' \ge t_{q-\Delta}$ such that $\theta^r_\Delta(t',q) \le \theta^\infty_{\Delta}(q) + \epsilon$. Since $\theta^r_\Delta(t',q) \ge \theta^\infty_{\Delta}(q) - \epsilon$, we now obtain that $\forall t \ge t_q, |\theta^r_\Delta(t,q) - \theta^\infty_{\Delta}(q)| \le \epsilon $.\\

Finally, if $\theta^r_\Delta(t_{q-\Delta},q) < \theta^\infty_{\Delta}(q) - \epsilon$, then there are two possibilities. The first one is that the function $t \ge t_{q-\Delta} \mapsto \theta^r_\Delta(t,q)$ is increasing and in that case it is bounded from above by $\theta^\infty_{\Delta}(q) + \epsilon$ and must converge. Since $\lim_{t \to +\infty} \theta_{\Delta}^r(t,q-\Delta) = \theta^\infty_{\Delta}(q-\Delta)$, the only possible limit for $\theta_{\Delta}^r(t,q)$ is $\theta^\infty_{\Delta}(q)$. The second possibility is that $t \ge t_{q-\Delta} \mapsto \theta^r_\Delta(t,q)$ is not an increasing function and in that case there must exists $t' \ge t_{q-\Delta}$ such that $\theta^r_\Delta(t',q) \ge \theta^\infty_{\Delta}(q) - \epsilon$. Since $\theta^r_\Delta(t',q) \le \theta^\infty_{\Delta}(q) + \epsilon$, we now obtain that $\forall t \ge t_q, |\theta^r_\Delta(t,q) - \theta^\infty_{\Delta}(q)| \le \epsilon $.\\

The conclusion is that $\limsup_{t \to +\infty} |\theta^r_\Delta(t,q) - \theta^\infty_{\Delta}(q)| \le \epsilon$. Sending $\epsilon$ to $0$, we get the result for $\theta_{\Delta}$.\\

The result for the optimal quote is then straightforward.\qed\\

\begin{figure}[!h]
\center
  \includegraphics[width=400pt]{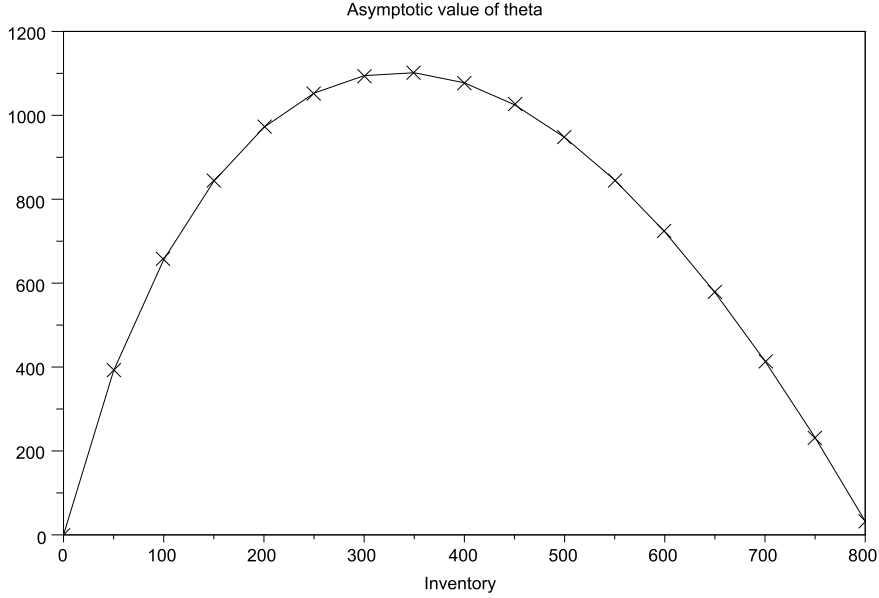}
  \caption{$\theta_{\Delta}^\infty(q)$ for $\Lambda_{\Delta}$ defined above, $q_0=800$, $\Delta = 50$, $\mu = 0$, $\sigma = 0.3 \mathrm{\;} (\mathrm{ Tick}.\mathrm{s}^{-\frac{1}{2}})$, $\gamma =  0.001\mathrm{\;}(\mathrm{Tick}^{-1})$.}
  \label{asymp}
\end{figure}

These asymptotic formulae deserve some comments. Firstly, regarding the above discussion on monotonicity, we know that $H_{\Delta}$ is a decreasing function. therefore, the asymptotic limit $\theta^\infty_{\Delta}(\cdot)$ is either a decreasing function (when $\frac 12 \gamma^2\sigma^2 \Delta^2 - \gamma \mu \Delta > H_\Delta(0)$) or a function that is first increasing and then decreasing (otherwise) -- see Figure \ref{asymp} in our case. Secondly, coming to the optimal quotes and the role of the parameters, we can analyze the way $\delta_\Delta^{*\infty}$ depends on $\mu$, $\sigma$, $\gamma$ and $\Lambda_{\Delta}$. The best way to proceed is to use the expression for $H_{\Delta}$ found in Lemma \ref{l1} and to notice that an equivalent way to define $\delta_\Delta^{*\infty}$ is through the following implicit characterization:

$$- \mu q_0 + \frac 12 \gamma \sigma^2 q_0^2 = \Delta \frac{\Lambda_{\Delta}(\delta_\Delta^{*\infty})^2}{\gamma\Delta \Lambda_{\Delta}(\delta_\Delta^{*\infty}) - \Lambda'_{\Delta}(\delta_\Delta^{*\infty})}.$$

It is then straightforward to see that $\delta_\Delta^{*\infty}$ is an increasing function of $\mu$. A trader expecting the stock price to go up is indeed encouraged to slow down the liquidation process. Similarly, we see that $\delta_\Delta^{*\infty}$ decreases as $\sigma$ increases. An increase in $\sigma$ corresponds to an increase in price risk and this provides the trader with an incentive to speed up the execution process. Therefore, it is natural that the asymptotic quote be a decreasing function of $\sigma$.\\
Differentiating the above expression with respect to $\gamma$, we see that the asymptotic quote decreases as the risk aversion increases. An increase in risk aversion forces indeed the trader to reduce both non-execution risk and price risk and this leads to posting orders with lower prices.\\
Now, if one replaces the intensity function $\Lambda_{\Delta}$ by $\lambda \Lambda_{\Delta}$ where $\lambda > 1$, then it results in an increase in $\delta_\Delta^{*\infty}$. This is natural because when the rate of arrival of liquidity-taking orders increases, the trader is more likely to liquidate his shares faster and posting deeper into the book allows for larger profits.\\

\subsection{The influence of $\Delta$}

In addition to the asymptotic regime, we can consider different sizes $\Delta$ of orders.

\begin{figure}[!h]
\center
  \includegraphics[width=400pt]{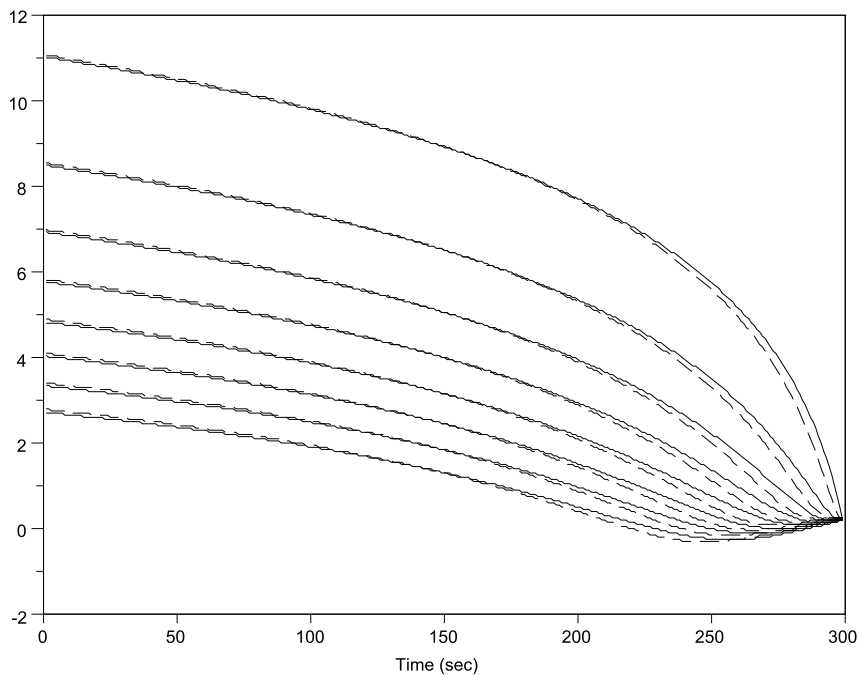}\\
  \caption{Optimal quotes for $q\in \lbrace 50, 100, \ldots, 400\rbrace$, for $q_0=600$, $T=1200 \mathrm{\;} (\mathrm{s})$, $\mu = 0$, $\sigma = 0.3 \mathrm{\;} (\mathrm{ Tick}.\mathrm{s}^{-\frac{1}{2}})$, $A = 0.1 \mathrm{\;} (\mathrm{s}^{-1})$, $k = 0.3 \mathrm{\;} (\mathrm{Tick}^{-1})$, $\gamma =  0.001\mathrm{\;}(\mathrm{Tick}^{-1})$ and $\ell(q) = \ell=3\mathrm{\;}(\mathrm{Tick})$. Line: $\Lambda(\delta) = A e^{-k\delta}$ and $\Delta = 50$. Dotted line: $\Lambda(\delta) = 2 A e^{-k\delta}$ and $\Delta = 25$. }
  \label{delta_Delta}
\end{figure}

We see on Figure \ref{delta_Delta}, that there is little difference between the two cases we considered. This is linked to the existence of a limit regime as $\Delta \to 0$ and the next section is dedicated to its analysis. The limiting equation for $\theta_\Delta$ will turn out to be a classical equation in optimal liquidation theory (see Section 5).

\section{Limit regime $\Delta \to 0$}

In the preceding sections, the size of the order posted by the trader was constant equal to $\Delta$, a size that is supposed to be small with respect to $q_0$. As a consequence, the question of the limiting behavior when $\Delta$ tends to $0$ is relevant.\footnote{Although this is not recalled, it is assumed that $\Delta$ is always chosen as a fraction of $q_0$.} To that purpose we need to make an assumption on the behavior of the intensity function with respect to the order size $\Delta$. The ``right'' scaling (already used above for the numerics underlying Figure \ref{delta_Delta}) is to suppose that

$$\Lambda_\Delta(\delta) = \frac{\Lambda(\delta)}{\Delta}.$$

With this scaling , along with additional technical hypotheses, our goal is to prove the following Theorem that is rather technical and echoes the results obtained by \cite{bayraktar2011liquidation}, here with risk aversion (\emph{i.e.} $\gamma > 0$) whereas \cite{bayraktar2011liquidation} deals with the risk-neutral case:

\begin{Theorem}[Limit regime $\Delta \to 0$]
\label{t3}
Let us suppose that:
\begin{itemize}
  \item $\Lambda(\delta) \Lambda''(\delta) < 2{\Lambda'}(\delta)^2$
  \item $\lim_{\delta \to +\infty} \delta \Lambda(\delta) = 0$
  \item $\ell$ is a continuous function
\end{itemize}

For a given $\Delta>0$, let us define $\theta^c_{\Delta}$ on $[0,T]\times[0,q_0]$ by:

$$\theta^c_{\Delta}(t,q) = \begin{cases} \theta_{\Delta}(t,0) , & \text{if } q=0,\\
                                        \theta_{\Delta}\left(t,(k+1)\Delta\right) , & \text{if } q \in (k\Delta, (k+1)\Delta].\\ \end{cases}$$

Then $\theta^c_{\Delta}$ converges uniformly toward a continuous function $\theta : [0,T]\times[0,q_0] \to \mathbb{R}$ that is the unique viscosity solution of the equation $(\textrm{HJ}_{\textrm{lim}})$:

$$ \begin{cases} -\gamma \partial_t \theta(t,q) - \gamma \mu q + \frac 12 \gamma^2 \sigma^2 q^2 -  H\left( \partial_q \theta(t,q) \right) = 0 , & \text{on } [0,T)\times(0,q_0],\\
\theta(t,q) = 0, & \text{on } [0,T]\times\lbrace0\rbrace,\\
\theta(t,q) = -\ell(q) q, & \text{on } \lbrace T\rbrace\times[0,q_0],\\
\end{cases}, \qquad (\textrm{HJ}_{\textrm{lim}})$$
where $H(p) = \gamma \sup_{\delta} \Lambda(\delta) (\delta - p)$ and where the terminal condition and the boundary condition are in fact satisfied is the classical sense.
\end{Theorem}

To prove this theorem, we first need to study $H$ and the convergence of the Hamiltonian functions $H_\Delta$ towards $H$. We start with a counterpart of Lemma \ref{l1} that requires $\Lambda(\delta) \Lambda''(\delta) < 2{\Lambda'}(\delta)^2$.

\begin{Lemma}
\label{l2}
Let us define $L(p,\delta) = \Lambda(\delta)\left( \delta - p\right)$.\\

$\forall p \in \mathbb{R}$, $\delta \mapsto L(p,\delta)$ attains its maximum at $\tilde{\delta}^*(p)$ uniquely characterized by:

$$\tilde{\delta}^*(p) + \frac{\Lambda (\tilde{\delta}^*(p))}{\Lambda'(\tilde{\delta}^*(p))} = p.$$

Moreover, $p \mapsto \tilde{\delta}^*(p)$ is a $C^1$ function.\\

Subsequently, $H$ is a $C^1$ function with:

$$H(p) = \gamma \frac{\Lambda(\tilde{\delta}^*(p))^2}{ - \Lambda'(\tilde{\delta}^*(p))}.$$
\end{Lemma}

\textbf{Proof:} The proof is similar to the proof of Lemma \ref{l1}.\\

Now, we can state a result about convergence that also provides a uniform bound for the hamiltonian functions:

\begin{Lemma}
\label{l3}
$H_{\Delta}$ converges locally uniformly towards $H$ when $\Delta \to 0$ with $\forall p \in \mathbb{R}, H_\Delta(p) \le H(p)$.
\end{Lemma}

\textbf{Proof:}\\

For a fixed $x \ge 0$, the function $f :\Delta \in \mathbb{R_+} \mapsto \frac{1-e^{-\Delta x}}{\Delta}$ is a decreasing function ($f(0) = x$).\\

Hence, $$\forall 0<\Delta'<\Delta,\quad \sup_{\delta \ge p} \Lambda(\delta) \frac{1-e^{-\gamma\Delta(\delta - p)}}{\Delta} \le  \sup_{\delta \ge p} \Lambda(\delta) \frac{1-e^{-\gamma \Delta'(\delta - p)}}{\Delta'} \le \gamma \sup_{\delta \ge p} \Lambda(\delta) (\delta - p).$$
This gives:
$$H_{\Delta}(p) \le H_{\Delta'}(p) \le H(p).$$

Now, because $H$ is continuous, using Dini's theorem, if we prove that convergence is pointwise, convergence will be locally uniform. We then only need to prove pointwise convergence of $H_\Delta$ toward $H$. Using Lemma \ref{l1} and Lemma \ref{l2}, we see that it is sufficient to prove that $\tilde{\delta}^*_{\Delta}$ converges pointwise towards $\tilde{\delta}^*$.

For that purpose, notice that the sequence of functions $f_{\Delta}(x) = x - \frac{1}{\gamma \Delta} \log\left( 1 - \gamma\Delta \frac{\Lambda(x)}{\Lambda' (x)}\right)$ is an increasing sequence of increasing functions. Hence, by the unique characterizations of $\tilde{\delta}^*_{\Delta}(p)$ and $\tilde{\delta}^*(p)$, we see that $\tilde{\delta}^*_{\Delta}(p)$ increases as $\Delta$ decreases to $0$ and is bounded from above by $\tilde{\delta}^*(p)$ which is the only possible limit. Hence $\tilde{\delta}^*_{\Delta}(p) \to \tilde{\delta}^*(p)$ as $\Delta \to 0$ and this proves the result.\qed\\

Now, we provide a uniform bound for the $\theta_{\Delta}$ that will be important in the proof of Theorem \ref{t3}.

\begin{Proposition}[Bounds for $\theta_{\Delta}$]
$\forall t\in [0,T], \forall q \in \lbrace 0, \Delta, \ldots , q_0 \rbrace$,
$$-\ell (q_0)q_0 - \mu^- q_0 (T-t) -\frac{1}{2} \gamma \sigma^2 q_0^2 (T-t) \le \theta_\Delta(t,q) \le \mu^+ q_0 (T-t) + \frac{1}{\gamma} H(0)(T-t).$$
\end{Proposition}

\textbf{Proof:}\\

To prove these inequalities, we use the comparison principle of Proposition \ref{comp}.\\

If $\overline{\theta}_{\Delta}(t,q) = \mu^+ q_0 (T-t) + \frac{1}{\gamma} H_{\Delta}(0) (T-t)$, then:
$$\overline{\theta}_{\Delta}(T,q) = 0 \geq -\ell(q) q, \qquad \overline{\theta}_{\Delta}(t,0) = \mu^+ q_0 (T-t) + \frac{1}{\gamma} H_{\Delta}(0)(T-t) \geq 0,$$
and
$$\gamma\partial_t \overline{\theta}_{\Delta}(t,q) + \gamma \mu q -
\frac{1}{2}\gamma^2\sigma^2 q^2
+ H_{\Delta}\left( \frac{\overline{\theta}_{\Delta}(t,q) - \overline{\theta}_{\Delta}(t,q-\Delta)}{\Delta} \right)
$$
$$
= -\gamma \frac{1}{\gamma} H_{\Delta}(0) -\gamma \mu^+ q_0 + \gamma \mu q  - \frac{1}{2}\gamma^2\sigma^2 q^2 + H_{\Delta}(0)$$$$ = -\gamma (\mu^+ q_0 - \mu q) - \frac{1}{2}\gamma^2\sigma^2 q^2 \le 0.
$$

Therefore, $\theta_\Delta(t,q) \le \overline{\theta}_{\Delta}(t,q) = \mu^+ q_0 (T-t) + \frac{1}{\gamma} H_{\Delta}(0)(T-t)$.\\

The uniform upper bound is then obtained using Lemma \ref{l3}.\\

Now, if $\underline{\theta}_{\Delta}(t,q) = -\ell(q_0) q_0 - \mu^- q_0 (T-t) - \frac{1}{2}\gamma\sigma^2 q_0^2 (T-t)$, then:
$$\underline{\theta}_{\Delta}(T,q) = - \ell(q_0) q_0 \leq -\ell(q) q, \qquad \underline{\theta}_{\Delta}(t,0) = -\ell(q_0) q_0 - \mu^- q_0 (T-t) - \frac{1}{2}\gamma^2\sigma^2 q_0^2 (T-t) \leq 0,$$
and
$$\gamma\partial_t \underline{\theta}_{\Delta}(t,q) + \gamma \mu q -
\frac{1}{2}\gamma^2\sigma^2 q^2
+ H_{\Delta}\left( \frac{\underline{\theta}_{\Delta}(t,q) - \underline{\theta}_{\Delta}(t,q-\Delta)}{\Delta} \right)
$$
$$
= \gamma \mu^- q_0 + \frac{1}{2}\gamma^2\sigma^2 q_0^2 + \gamma \mu q  - \frac{1}{2}\gamma^2\sigma^2 q^2 + H_{\Delta}(0)$$
$$ \ge \gamma (\mu^- q_0 + \mu q) +  \frac{1}{2}\gamma^2\sigma^2 (q_0^2 - q^2) \ge 0.
$$

Therefore, $\theta_\Delta(t,q) \ge \underline{\theta}_{\Delta}(t,q) = -\ell(q_0) q_0 - \mu^- q_0 (T-t) - \frac{1}{2}\gamma\sigma^2 q_0^2 (T-t)$.\qed\\

We are now ready to start the proof of Theorem \ref{t3}.\\

\textbf{Proof of Theorem \ref{t3}:}\\

We first introduce the following half-relaxed limit functions:

$$\overline{\theta}(t,q) = \limsup_{j \to +\infty} \limsup_{\Delta \to 0} \sup\left\lbrace \theta^c_{\Delta}(t',q'), \quad |t'-t|+|q'-q| \le \frac 1j\right\rbrace,$$

$$\underline{\theta}(t,q) = \liminf_{j \to +\infty} \liminf_{\Delta \to 0} \inf\left\lbrace \theta^c_{\Delta}(t',q'), \quad |t'-t|+|q'-q| \le \frac 1j\right\rbrace.$$

$\overline{\theta}$ and $\underline{\theta}$ are respectively upper semi-continuous and lower semi-continuous and the goal of the proof is to show that they are equal to one another and solution of the partial differential equation $(\textrm{HJ}_{\textrm{lim}})$.\\

\emph{Step 1:} $\overline{\theta}$ and $\underline{\theta}$ are respectively viscosity subsolution and viscosity supersolution of the equation $(\textrm{HJ}_{\textrm{lim}})$.\\

To prove this point, let us consider $(t^*,q^*)$ and a test function $\phi$ such that $\overline{\theta} - \phi$ attains a local maximum at $(t^*,q^*)$. Without loss of generality, we can assume that $\phi(t^*,q^*) = \overline{\theta}(t^*,q^*)$ and consider $r>0$ such that:
\begin{itemize}
  \item the maximum is global on the ball of radius $r$ centered in $(t^*,q^*)$.
  \item outside of this ball, $\phi \ge 2 \sup_{\Delta} \|\theta_{\Delta}\|_{\infty}$ -- this value being finite because of the uniform bound obtained in the above Proposition.\\
\end{itemize}

Following Barles-Souganidis methodology \cite{barles1990convergence}, we know that there exists a sequence $(\Delta_n,t_n,q_n)_n$ such that:
\begin{itemize}
  \item $\Delta_n \to 0$, $(t_n,q_n) \to (t^*,q^*)$
  \item $\theta^c_{\Delta_n}(t_n,q_n) \to \overline{\theta}(t^*,q^*)$
  \item $\theta^c_{\Delta_n} - \phi$ has a global maximum at $(t_n,q_n)$.
\end{itemize}

Now, because of the definition of $\theta^c_{\Delta_n}$, if $(t^*,q^*) \in [0,T)\times(0,q_0]$ we can always suppose that $q_n \ge \Delta_n$ and $t_n \neq T$ and then, by definition of $\theta_{\Delta_n}$:

$$-\gamma \partial_t \theta_{\Delta_n}(t_n,q_n) - \gamma \mu q_n +
\frac{1}{2} \gamma^2 \sigma^2 q_n^2
-  H_{\Delta_n}\left( \frac{\theta_{\Delta_n}(t_n,q_n) - \theta_{\Delta_n}(t_n,q_n-\Delta_n)}{\Delta_n} \right) = 0.
$$

Because $H_{\Delta_n}$ is decreasing we have, by definition of $(t_n,q_n)$:

$$-H_{\Delta_n}\left( \frac{\theta_{\Delta_n}(t_n,q_n) - \theta_{\Delta_n}(t_n,q_n-\Delta_n)}{\Delta_n} \right) \ge -H_{\Delta_n}\left( \frac{\phi(t_n,q_n) - \phi(t_n,q_n-\Delta_n)}{\Delta_n} \right).$$

Similarly, since $t_n < T$, we have, for $h$ sufficiently small:
$$\theta_{\Delta_n}(t_n+h,q_n) - \theta_{\Delta_n}(t_n,q_n) \le \phi(t_n+h,q_n) - \phi(t_n,q_n).$$
Hence:
$$\partial_t \theta_{\Delta_n}(t_n,q_n) \le \partial_t \phi(t_n,q_n).$$

These inequalities give:

$$-\gamma \partial_t \phi(t_n,q_n) - \gamma \mu q_n +
\frac{1}{2} \gamma^2 \sigma^2 q_n^2
-  H_{\Delta_n}\left( \frac{\phi(t_n,q_n) - \phi(t_n,q_n-\Delta_n)}{\Delta_n} \right) \le 0.
$$

Using now the convergence of $(t_n,q_n)$ towards $(t^*,q^*)$ and the local uniform convergence of $H_{\Delta_n}$ towards $H$, we eventually obtain the desired inequality:

$$-\gamma \partial_t \phi(t^*,q^*) - \gamma \mu q^* +
\frac{1}{2} \gamma^2 \sigma^2 {q^*}^2
-  H\left( \partial_q\phi(t^*,q^*) \right) \le 0.
$$

We see that the boundaries corresponding to $q=q_0$ and $t=0$ play no role. However, we need to consider the cases $t^*=T$ and $q^*=0$.\\

If $t^*=T$ and $q^*\neq 0$ then there are two cases. If there are infinitely many indices $n$ such that $t_n < T$ then the preceding proof still works. Otherwise, for all $n$ sufficiently large, $\theta_{\Delta_n}(t_n,q_n) = -\ell(q_n) q_n$ and hence, passing to the limit, $\overline{\theta}(t^*,q^*) = -\ell(q^*) q^*$.\\

Eventually, we indeed have that:

$$\min(-\gamma \partial_t \phi(T,q^*) - \gamma \mu q^* +
\frac{1}{2} \gamma^2 \sigma^2 {q^*}^2
-  H\left( \partial_q\phi(T,q^*) \right), \overline{\theta}(T,q^*) + \ell(q^*) q^* ) \le 0.$$

If $q^*=0$ then there are also two cases. If there are infinitely many indices $n$ such that $q_n^* > 0$ and $t_n < T$ then the initial proof still works. Otherwise, for $n$ sufficiently large $\theta_{\Delta_n}(t_n,q_n) = -\ell(q_n) q_n$ or $\theta_{\Delta_n}(t_n,q_n) = 0$ and hence, passing to the limit, we obtain $\overline{\theta}(t^*,q^*) = 0$.\\

Eventually, we indeed have that:

$$\min(-\gamma \partial_t \phi(t^*,0)
-  H\left( \partial_q\phi(t^*,0) \right), \overline{\theta}(t^*,0)) \le 0.$$

We have proved that $\overline{\theta}$ is a subsolution of the equation in the viscosity and one can similarly prove that $\underline{\theta}$ is a supersolution.\\

\emph{Step 2:} $\forall q \in [0,q_0], \quad \overline{\theta}(T,q) = \underline{\theta}(T,q) = -\ell(q)q$.\\

We consider the test function $\phi(t,q) = C_{\epsilon}(T-t) + \frac{1}{\epsilon} (q-q_{ref})^2$ where $q_{ref} \in [0,q_0]$ is fixed, where $\epsilon>0$ is a constant and where $C_\epsilon$ is a constant that depends on $\epsilon$, and that will be fixed later.\\

Let $(t_\epsilon,q_\epsilon)$ be a maximum point of $\overline{\theta} - \phi$ on $[0,T]\times[0,q_0]$. Then:
$$\overline{\theta}(T,q_{ref}) \le \overline{\theta}(t_\epsilon,q_\epsilon) - C_{\epsilon}(T-t_\epsilon) - \frac{1}{\epsilon} (q_\epsilon-q_{ref})^2.$$

This inequality gives $q_\epsilon \to q_{ref}$ as $\epsilon \to 0$.\\

Now,
$$-\gamma \partial_t \phi(t_\epsilon,q_\epsilon) - \gamma \mu q_\epsilon +
\frac{1}{2} \gamma^2 \sigma^2 {q_\epsilon}^2
-  H\left( \partial_q\phi(t_\epsilon,q_\epsilon) \right) = \gamma C_\epsilon - \gamma \mu q_\epsilon + \frac{1}{2} \gamma^2 \sigma^2 {q_\epsilon}^2 - H\left(\frac{2}{\epsilon} (q_\epsilon-q_{ref})\right)  $$
$$\ge \gamma C_\epsilon - H\left(-\frac{2q_0}{\epsilon}\right)+ \inf_{q \in [0,q_0]} \left(-\gamma \mu q + \frac 12 \gamma^2 \sigma^2 q^2\right).$$
Hence, if $ C_\epsilon$ is positive and greater than $\frac{1}{\gamma} H\left(-\frac{2q_0}{\epsilon}\right) -\inf_{q \in [0,q_0]} \left(-\gamma \mu q + \frac 12 \gamma^2 \sigma^2 q^2\right) + 1$, we see that we must have either $t_\epsilon = T$ and $\overline{\theta}(t_\epsilon,q_\epsilon) \le -\ell(q_\epsilon) q_\epsilon$ or (for sufficiently small $\epsilon$, only in the case where $q_{ref}=0$), $\overline{\theta}(t_\epsilon,q_\epsilon) \le 0$.\\

If $q_{ref} \neq 0$, we then write, for $\epsilon$ sufficiently small:

$$\overline{\theta}(T,q_{ref}) \le \overline{\theta}(t_\epsilon,q_\epsilon) - C_{\epsilon}(T-t_\epsilon) - \frac{1}{\epsilon} (q_\epsilon-q_{ref})^2 \le \overline{\theta}(t_\epsilon,q_\epsilon) \le -\ell(q_\epsilon) q_\epsilon.$$

Sending $\epsilon$ to $0$ we obtain $$\overline{\theta}(T,q_{ref}) \le -\ell(q_{ref}) q_{ref}.$$

If $q_{ref} = 0$, then we know from the above two inequalities that:
$$\overline{\theta}(T,q_{ref}) \le \overline{\theta}(t_\epsilon,q_\epsilon) - C_{\epsilon}(T-t_\epsilon) - \frac{1}{\epsilon} (q_\epsilon-q_{ref})^2 \le \overline{\theta}(t_\epsilon,q_\epsilon) \le \max(-\ell(q_\epsilon) q_\epsilon,0) = 0.$$

Hence, $\forall q \in [0,q_0], \overline{\theta}(T,q) \le -\ell(q)q$ and the same proof works for the supersolution\footnote{The only difference is that we have to pass to the limit in the case $q_{ref} = 0$ to obtain the conclusion.} to get $\forall q \in [0,q_0], \underline{\theta}(T,q) \ge -\ell(q)q$ .\\

As a consequence $\forall q \in [0,q_0], \overline{\theta}(T,q) \le -\ell(q) q \le \underline{\theta}(T,q)$, and eventually, because $\underline{\theta}(T,q) \le \overline{\theta}(T,q)$, we obtain that $\forall q \in [0,q_0], \overline{\theta}(T,q) = -\ell(q) q = \underline{\theta}(T,q)$.\\

\emph{Step 3:} $\forall t \in [0,T], \quad \underline{\theta}(t,0)  =0$\\

Concerning the boundary condition corresponding to $q=0$ we can apply the same ideas but only to the supersolution $\underline{\theta}$.\\

Let us consider indeed $t_{ref} \in [0,T)$ and the test function $\phi(t,q) = -C_{\epsilon}q - \frac{1}{\epsilon} (t-t_{ref})^2$. Then, let $(t_\epsilon,q_\epsilon)$ be a minimum point of $\underline{\theta} - \phi$ on $[0,T]\times[0,q_0]$. We have:
$$\underline{\theta}(t_{ref},0) \ge \underline{\theta}(t_\epsilon,q_\epsilon) +C_{\epsilon}q_\epsilon + \frac{1}{\epsilon} (t_\epsilon-t_{ref})^2.$$

This inequality gives $t_\epsilon \to t_{ref}$ as $\epsilon \to 0$.

Now,

$$-\gamma \partial_t \phi(t_\epsilon,q_\epsilon) - \gamma \mu q_\epsilon +
\frac{1}{2} \gamma^2 \sigma^2 {q_\epsilon}^2
-  H\left( \partial_q\phi(t_\epsilon,q_\epsilon) \right) = 2\gamma \frac{t-t_{ref}}{\epsilon}- \gamma \mu q_\epsilon + \frac{1}{2} \gamma^2 \sigma^2 {q_\epsilon}^2 - H\left(-C_\epsilon\right)  $$
$$\le 2 \gamma \frac{T}{\epsilon} - \gamma \mu q_0 + \frac{1}{2} \gamma^2 \sigma^2 {q_0}^2 - H\left(-C_\epsilon\right).$$
Since $\lim_{p\to -\infty} H(p) = +\infty$, we can always choose $C_\epsilon \ge 0$ so that the above expression is strictly negative.\\

As a consequence, for $\epsilon$ sufficiently small, we must have $q_{\epsilon} = 0$ and $\underline{\theta}(t_\epsilon,0) \ge 0$.\\

Consequently:

$$\underline{\theta}(t_{ref},0) \ge \underline{\theta}(t_\epsilon,q_\epsilon) +C_{\epsilon}q_\epsilon + \frac{1}{\epsilon} (t_\epsilon-t_{ref})^2 \ge 0.$$

This result being already true for $t_{ref} = T$, we have that $\underline{\theta}(t,0) \ge 0, \forall t \in [0,T]$ and in fact $\underline{\theta}(t,0) = 0, \forall t \in [0,T]$ because of the definition of $\underline{\theta}(t,0)$.\\

\emph{Step 4:} $\forall t \in [0,T]$, there exists a sequence $(t_n,q_n)_n$ such that $t_n \neq t$, $q_n\neq 0$, $(t_n,q_n) \to (t,0)$, and $\underline{\theta}(t_n,q_n) \to 0$\\

To prove this claim, we prove that $g_{\Delta}(t) = \theta_{\Delta}(T-t,\Delta)$ converges uniformly (in $t$) toward $0$ as $\Delta \to 0$.\\

By definition, $g_{\Delta}(0) = - \ell(\Delta) \Delta$ and  $g'_{\Delta}(t) = \mu \Delta -\frac{1}{2} \gamma \sigma^2 \Delta^2 + \frac 1{\gamma} H_\Delta\left(\frac{g_{\Delta}(t)}{\Delta}\right)$.\\

We now distinguish two cases:\\

Case 1: $\mu \le 0$.\\

The stationary state of the above ODE is $g_{\Delta}^{\infty} = \Delta H_{\Delta}^{-1}(\frac{1}{2} \gamma^2 \sigma^2 \Delta^2 - \gamma \mu \Delta)$ and $g_\Delta$ is increasing on $\lbrace g_\Delta \le g_{\Delta}^{\infty} \rbrace$. Since, $\Delta H_{\Delta}^{-1}(\frac{1}{2} \gamma^2 \sigma^2 \Delta^2- \gamma \mu \Delta) \ge \Delta H_{\Delta'}^{-1}(\frac{1}{2} \gamma^2 \sigma^2 \Delta^2 - \gamma \mu \Delta)$ as soon as $\Delta < \Delta'$, $g_{\Delta}^{\infty}$ is positive for $\Delta$ sufficiently small. As a consequence, since $g_{\Delta}(0) \le 0$, $g_{\Delta}$ is increasing on [0,T].\\

Now, $g'_{\Delta}(t) \le  \frac 1{\gamma} H\left(\frac{g_{\Delta}(t)}{\Delta}\right)$. Hence,

$$\int_0^{g_\Delta(t)} \frac{\gamma}{H\left(\frac{y}{\Delta}\right)} dy \le \int_{-\ell(\Delta)\Delta}^{g_\Delta(t)} \frac{\gamma}{H\left(\frac{y}{\Delta}\right)} dy \le t,$$
and this gives $g_{\Delta}(t) \le \Delta G^{-1}\left(\frac{T}{\gamma \Delta}\right)$, where $G(x) = \int_{0}^x \frac 1{H(y)}dy$.\\

Consequently,

$$\forall t \in [0,T], \quad -\ell(\Delta)\Delta \le g_{\Delta}(t) \le \Delta G^{-1}\left(\frac{T}{\gamma \Delta}\right).$$
Since $\lim_{\delta \to +\infty} \delta \Lambda(\delta) = 0$, we have $\lim_{x \to +\infty} H(x) = 0$, and therefore $\lim_{x \to +\infty} \frac{G(x)}{x} = +\infty$. This proves that $g_{\Delta}$ converges uniformly toward $0$ on $[0,T]$.

Case 2: $\mu > 0$.\\

In this second case, we know that $\exists \Delta_0$ such that $\Delta \in (0, \Delta_0) \mapsto \mu \Delta - \frac{1}{2} \gamma \sigma^2 \Delta^2$ is an increasing and positive function. Now, let us introduce $\Delta_1 \in (0, \Delta_0)$. $\forall \Delta \in (0, \Delta_1)$, we have:

$$\int_0^{g_\Delta(T)} \frac{\gamma}{\gamma \mu \Delta_1 -\frac{1}{2} \gamma^2 \sigma^2 \Delta_1^2 + H\left(\frac{y}{\Delta}\right)} dy$$
 $$\le \int_{-\ell(\Delta)\Delta}^{g_\Delta(T)}   \frac{\gamma}{ \gamma \mu \Delta -\frac{1}{2} \gamma^2 \sigma^2 \Delta^2+H_\Delta\left(\frac{y}{\Delta}\right)} dy = T.$$

This gives $g_{\Delta}(T) \le \Delta G_{\Delta_1}^{-1}\left(\frac{T}{\gamma \Delta}\right)$, where $G_{\Delta_1}(x) = \int_{0}^x \frac 1{\gamma \mu \Delta_1 -\frac{1}{2} \gamma^2 \sigma^2 \Delta_1^2 + H(y)}dy$.

Now, because $\lim_{x \to +\infty} H(x) = 0$, we have:
$$\liminf_{x \to  +\infty} \frac{G_{\Delta_1}(x)}{x} \ge \frac{1}{\gamma \mu \Delta_1 -\frac{1}{2} \gamma^2 \sigma^2 \Delta_1^2}.$$
Hence, $$\liminf_{\Delta \to 0} \frac{T}{\gamma \Delta G_{\Delta_1}^{-1}\left(\frac{T}{\gamma \Delta}\right)} \ge \frac{1}{\gamma \mu \Delta_1 -\frac{1}{2} \gamma^2 \sigma^2 \Delta_1^2}.$$
This gives $$\limsup_{\Delta \to 0} g_{\Delta}(T) \le T (\mu \Delta_1 -\frac{1}{2} \gamma \sigma^2 \Delta_1^2),$$ and sending $\Delta_1$ to $0$, we get:
$$\limsup_{\Delta \to 0} g_{\Delta}(T) \le 0.$$ The result is then proved since:

$$\forall t \in [0,T],\quad -\ell(\Delta)\Delta \le g_{\Delta}(t) \le g_{\Delta}(T).$$

\emph{Step 5:} Comparison principle: $\overline{\theta} \le \underline{\theta}$.\\

Let us consider $\alpha > 0$ and the maximum $M = \max_{(t,q) \in [0,T]\times[0,q_0]}\overline{\theta}(t,q) - \underline{\theta}(t,q) - \alpha (T-t)$. If $ m = \max_{t \in [0,T]}\overline{\theta}(t,0) - \underline{\theta}(t,0) - \alpha (T-t) < M$ then we distinguish two cases.\\

Case 1: $m \le 0$.\\

We introduce $\Phi_{\epsilon}(t,q,t',q') = \overline{\theta}(t,q) - \underline{\theta}(t',q') - \alpha(T-t) - \frac{(q-q')^2}{\epsilon} - \frac{(t-t')^2}{\epsilon}$.\\

Let us consider $(t_\epsilon,q_\epsilon,t'_\epsilon,q'_\epsilon)$ a maximum point of $\Phi_{\epsilon}$. We have $M \le \Phi_{\epsilon}(t_\epsilon,q_\epsilon,t'_\epsilon,q'_\epsilon)$ and we are going to prove that $\liminf_{\epsilon\to 0} \Phi_{\epsilon}(t_\epsilon,q_\epsilon,t'_\epsilon,q'_\epsilon) \le 0$.\\

We have $\Phi_{\epsilon}(t_\epsilon,q_\epsilon,t_\epsilon,q_\epsilon) \le \Phi_{\epsilon}(t_\epsilon,q_\epsilon,t'_\epsilon,q'_\epsilon)$ and hence $\frac{(q_\epsilon - q'_\epsilon)^2}{\epsilon} + \frac{(t_\epsilon-t'_\epsilon)^2}{\epsilon}$ is bounded. As a consequence, $ t_\epsilon-t'_\epsilon \to 0$ and $q_\epsilon - q'_\epsilon \to 0$.\\

Now, if for all $\epsilon$ sufficiently small we have $(t_\epsilon,q_\epsilon,t'_\epsilon,q'_\epsilon) \in [0,T)\times(0,q_0]\times[0,T)\times(0,q_0]$ then we have:

$$-2\gamma\frac{t_\epsilon- t'_\epsilon}{\epsilon} + \gamma\alpha - \gamma \mu q_\epsilon  + \frac 12 \gamma^2 \sigma^2 q_\epsilon^2 - H(2\frac{q_\epsilon- q'_\epsilon}{\epsilon}) \le 0,$$

and

$$-2\gamma\frac{t_\epsilon- t'_\epsilon}{\epsilon} - \gamma \mu q'_\epsilon + \frac 12 \gamma^2 \sigma^2 {q'_\epsilon}^2 - H(2\frac{q_\epsilon- q'_\epsilon}{\epsilon}) \ge 0.$$

Hence $ \gamma\alpha -\gamma \mu (q_\epsilon - q'_\epsilon) + \frac 12 \gamma^2 \sigma^2 (q_\epsilon^2 - {q'_\epsilon}^2) \le 0$ and this is a contradiction as we send $\epsilon$ to 0.\\

The consequence is that there exists a sequence $\epsilon_n \to 0$ such that $(t_{\epsilon_n},q_{\epsilon_n},t'_{\epsilon_n},q'_{\epsilon_n})$ verifies $t_{\epsilon_n}=T, \forall n$ or $q_{\epsilon_n} = 0, \forall n$ or $t'_{\epsilon_n}, \forall n$ or $q'_{\epsilon_n}, \forall n$.\\

Then
$$\limsup_{n \to +\infty} \Phi_{\epsilon_n}(t_{\epsilon_n},q_{\epsilon_n},t'_{\epsilon_n},q'_{\epsilon_n}) \le \limsup_{n \to +\infty} \overline{\theta}(t_{\epsilon_n},q_{\epsilon_n}) - \underline{\theta}(t'_{\epsilon_n},q'_{\epsilon_n}) - \alpha(T-t_{\epsilon_n})$$$$ \le \max_{(t,q) \in (\lbrace T\rbrace \times [0,q_0]) \cup ([0,T]\times \rbrace q_0\rbrace)} \overline{\theta}(t,q) - \underline{\theta}(t,q) - \alpha (T-t) \le \max(0,m) \le 0.$$

Hence in that case $M \le 0$.\\

Case 2: $m > 0$.\\

In that case, we replace $\underline{\theta}$ by $\underline{\theta} + m$ in the above case, and we obtain, instead of $M\le0$, the inequality $M\le m$ which contradicts our hypothesis.\\

It remains to consider the case $m=M$. We know then that the maximum $M$ is attained at a point $(t_{max},0)$ and we suppose that $M>0$.\\

From Step 4, we consider a sequence $(t_n,q_n)$ such that $t_n \neq t_{max}$, $q_n\neq 0$, $(t_n,q_n) \to (t_{max},0)$ and $\underline{\theta}(t_n,q_n) \to 0$.\\

We define: $$\Psi_n(t,q,t',q') = \overline{\theta}(t,q) - \underline{\theta}(t',q') - \alpha(T-t) - \frac{(t-t')^2}{|t_n - t_{max}|} - \left(\frac{q'-q}{q_n}-1\right)^2.$$

This function attains its maximum at a point $(t^*_n,q^*_n,{t'}^*_n,{q'}^*_n)$. We first consider the inequality $\Psi_n(t_{max},0,t_n,q_{max}) \le \Psi_n(t^*_n,q^*_n,{t'}^*_n,{q'}^*_n)$:

$$\overline{\theta}(t_{max},0) - \underline{\theta}(t_n,q_n) - \alpha(T-t_{max}) - |t_n - t_{max}|$$$$ \le \overline{\theta}(t^*_n,q^*_n) - \underline{\theta}({t'}^*_n,{q'}^*_n) - \alpha(T-t^*_n) - \frac{(t^*_n-{t'}^*_n)^2}{|t_n - t_{max}|} - \left(\frac{{q'}^*_n-q^*_n}{q_n}-1\right)^2.$$

We then have $t^*_n-{t'}^*_n \to 0$ and ${q'}^*_n-q^*_n \to 0$.\\

Hence $\limsup_n \overline{\theta}(t^*_n,q^*_n) - \underline{\theta}({t'}^*_n,{q'}^*_n) - \alpha(T-t^*_n) \le M$. Now, the maximum $M$ is also given by $\lim_n \overline{\theta}(t_{max},0) - \underline{\theta}(t_n,q_n) - \alpha(T-t_{max})$, and we obtain that the penalization term $\left(\frac{{q'}^*_n-q^*_n}{q_n}-1\right)^2$ converge to $0$.\\

This gives ${q'}^*_n = q^*_n + q_n + o(q_n) > 0$.\\

Now, if we have infinitely many $n$ such that ${t'}^*_n=T$, then:

$$M = \lim_n \overline{\theta}(t_{max},0) - \underline{\theta}(t_n,q_n) - \alpha(T-t_{max}) \le \sup_{q \in [0,q_0]} \overline{\theta}(T,q) - \underline{\theta}(T,q) = 0,$$
and this contradicts our hypothesis.\\

Otherwise, for all $n$ sufficiently large:

$$-2\gamma\frac{t^*_n- {t'}^*_n}{|t_n - t_{max}|} - \gamma \mu {q'}^*_n  + \frac 12 \gamma^2 \sigma^2 {{q'}^*_n}^2 - H\left(-\frac 2{q_n}\left({\frac{{q'}^*_n - q^*_n}{q_n}- 1 }\right)\right) \ge 0.$$

Now, going to the subsolution, if there are infinitely many $n$ such that $t^*_n=T$, then:

$$M = \lim_n \overline{\theta}(t_{max},0) - \underline{\theta}(t_n,q_n) - \alpha(T-t_{max}) \le \sup_{q \in [0,q_0]} \overline{\theta}(T,q) - \underline{\theta}(T,q) = 0,$$
in contradiction with our hypothesis.

Else, if $q^*_n = 0$ and $\overline{\theta}(t^*_n,q^*_n) \le 0$ for infinitely many $n$, then we obtain $M=\lim_n \overline{\theta}(t_{max},0) - \underline{\theta}(t_n,q_n) - \alpha(T-t_{max}) - |t_n - t_{max}| \le 0$ straightforwardly, and this contradicts our hypothesis. Hence, the viscosity inequality must be satisfied and we get:

$$-2\gamma\frac{t^*_n- {t'}^*_n}{|t_n - t_{max}|}  + \gamma \alpha - \gamma \mu {q}^*_n + \frac 12 \gamma^2 \sigma^2 {{q}^*_n}^2 - H\left(-\frac 2{q_n}\left({\frac{{q'}^*_n - q^*_n}{q_n}- 1 }\right)\right) \le 0.$$

Combining the two inequalities eventually leads to $\alpha \gamma \le 0$ as $n\to \infty$ and this is a contradiction.\\

We have obtained that $M\le 0$ and hence $\overline{\theta} - \underline{\theta} \le \alpha T$. Sending $\alpha$ to $0$, we get $\overline{\theta} \le \underline{\theta}$.\\

This proves that $\overline{\theta}=\underline{\theta}$ is in fact a continuous function that we call $\theta$, solution of the PDE $(\textrm{HJ}_{\textrm{lim}})$. Using the same techniques as above, it is clear that $\theta$ is the unique viscosity solution of $(\textrm{HJ}_{\textrm{lim}})$.\\

We have: $$\theta(t,q) = \underline{\theta}(t,q) \le \liminf_{\Delta \to 0} \theta_\Delta^c(t,q) \le \limsup_{\Delta \to 0} \theta_\Delta^c(t,q) \le \overline{\theta}(t,q) = \theta(t,q).$$

Hence $\theta(t,q) = \lim_{\Delta \to 0} \theta_\Delta^c(t,q)$ and, by the same token, $\lim_{\Delta \to 0, (t',q') \to (t,q) } \theta_\Delta^c(t',q') = \theta(t,q)$ so that the convergence is locally uniform and then uniform on the compact set $[0,T]\times[0,q_0]$.\qed\\

\section{Link with Almgren-Chriss}

\subsection{Interpretation of the PDE in the limit regime}

In the above section we proved that $\theta_\Delta$ converged to $\theta$, which is the unique continuous viscosity solution of $(\textrm{HJ}_{\textrm{lim}})$, the limit condition and boundary condition being satisfied in the classical sense.\\

Now, we are going to link this equation to a classical equation of Almgren-Chriss-like models. The intuition behind the link between our framework in the limit regime $\Delta \to 0$ and the Almgren-Chriss framework is that non-execution risk vanishes as $\Delta$ tends to $0$. Hence, the only remaining risk is price risk, corresponding to the term $\frac 12 \gamma^2 \sigma^2 q^2$ in the above equation. To see more precisely the correspondence between the two approaches, let us write the hamiltonian function as:

\begin{eqnarray*}
  H(p) &=& \gamma \sup_{\delta} \Lambda(\delta) (\delta - p) \\
   &=& \gamma \sup_{v > 0} v (\Lambda^{-1}(v) - p) \\
   &=& \gamma \sup_{v \ge 0} v\Lambda^{-1}(v) - pv,
\end{eqnarray*}
where the last equality holds since $\lim_{\delta \to +\infty} \delta \Lambda(\delta) = 0$.\\

Hence, if we define for $v > 0$, $f(v) = - \Lambda^{-1}(v)$, then we can define the function $\tilde{H}(p) = \sup_{v \ge 0} -f(v)v - pv$ and write the partial differential equation for $\theta$ as:

$$\begin{cases} - \partial_t \theta(t,q) - \mu q + \frac 12 \gamma \sigma^2 q^2 -  \tilde{H}\left( \partial_q \theta(t,q) \right) = 0 , & \text{on } [0,T)\times(0,q_0],\\
\theta(t,q) = 0, & \text{on } [0,T]\times\lbrace0\rbrace,\\
\theta(t,q) = -\ell(q) q, & \text{on } \lbrace T\rbrace\times[0,q_0].\\
\end{cases}$$

This equation is the Hamilton-Jacobi equation associated to the Almgren-Chriss optimal liquidation problem with an instantaneous market impact function (per share) $f$ and with a final discount $\ell(q_T)$ per share. More precisely, the above Hamilton-Jacobi equation is the Hamilton-Jacobi equation associated to the optimization problem:\footnote{In the usual Almgren-Chriss framework, liquidation is mandatory but the theory can easily be adapted to allow for a penalization term.}

$$\inf_{q \in AC(0,T), q(0) = q_0} \int_0^T \left({-q'(t) f(-q'(t)) - \mu q(t) + \frac 12 \gamma \sigma^2 q(t)^2}\right) dt + q(T) \ell(q(T)),$$ where $AC(0,T)$ is the set of absolutely continuous functions on $[0,T]$.\\

However, the instantaneous market impact function $f$ has here a rather unusual form since $f(v)$ is negative for $v < \Lambda(0)$ and positive for $v > \Lambda(0)$, whereas it is usually a positive function. This must be interpreted in a very simple way: if one needs to obtain an instantaneous volume lesser than $\Lambda(0)$, then one will choose a positive $\delta$. In other words, he will post a classical limit order, since we assumed that the reference price is the first bid quote -- this makes sense since non-execution risk disappears in the limit regime $\Delta \to 0$. On the contrary, if one needs an instantaneous volume greater than $\Lambda(0)$, then one will rely on a marketable limit order ($\delta < 0$).\\

The above discussion only makes sense at the limit, when non-execution risk does not exist anymore. However, it clarifies the meaning of negative $\delta$s. In particular, the above correspondence between our model and a model \emph{à la} Almgren-Chriss provides a possible way to solve one of the main practical problems of the model discussed in \cite{GLFT}: the interpretation of the intensity functions for negative values of $\delta$.

\subsection{Discussion on the choice of $\Lambda$}

The model we discuss in this paper does not consider explicitly market orders or limit orders but rather considers that there is, for each price $s^a = s+ \delta$, an instantaneous probability to obtain a trade at that price. In practice, this interpretation is perfectly suited to classical limit orders, but we need to provide an interpretation for the intensity function $\Lambda$ on the entire real line. In practice, since the model has been designed to liquidate a position with limit orders, it should not be used if the liquidation evidently requires liquidity-taking orders from the very beginning. However, it may happen, because of a slow execution, that the optimal quote in the model turns out to be negative\footnote{Although quotes evolve continuously between execution times, using the model in practice requires to post orders and to keep them in the order book for some time. Hence, the optimal quote at some point may in practice be strictly negative.} after some time.\\

This issue of negative $\delta$ was present in the model with exponential intensity functions introduced in \cite{GLFT}. Although the exponential form was justified for many stocks for positive $\delta$s, the intensity function was also of exponential form for $\delta < 0$, and this choice was dictated by mathematical needs rather than by empirical rationale. Since $\Lambda$ (or in practice $\Lambda_\Delta$) can be chosen in our setting, we can improve the initial model.\\
First, using statistics on execution, we can estimate the probability to be executed at any positive distance from the first bid limit. In practice the profile of the empirical intensity for positive $\delta$ is decreasing and may not be convex, especially when the bid-ask spread is large (this is the rationale underlying our choice -- for $\delta \ge 0$ -- on Figure \ref{lambda}). Then, once the function $\Lambda_{\Delta}$ has been calibrated for positive $\delta$, several natural choices are possible for $\Lambda_{\Delta}(\delta)$ when $\delta\le0$. Instead of extending the function for negative $\delta$ using a specific functional form as in \cite{GLFT}, we can assign to $\Lambda_\Delta(\delta)$ a constant value when $\delta \le 0$ as above for $\tilde{\Lambda}_\Delta$. This corresponds to a very conservative choice that basically prevents the use of marketable limit orders since, intuitively, the optimal quote will then always be positive. Another choice consists in using the parallel made between the usual literature and our framework in the limit regime $\Delta \to 0$. If we indeed omit non-execution risk, we can consider, for $\delta \le 0$, that $\Lambda_{\Delta}(\delta) = \frac{1}{\Delta} {f}^{-1}(-\delta)= \frac{1}{\Delta} \sup \lbrace v \ge 0, f(v) \le -\delta \rbrace$ where $v \mapsto f(v)$ is an instantaneous market impact function (average execution cost per share) that is typically equal to nought for small values of $v$ and increasing after a certain threshold.\\
This choice for $\Lambda_{\Delta}$ is however subject to several comments. First, the function $\Lambda_{\Delta}$ must satisfy the hypotheses of the model. In particular, it must be decreasing. However, it may happen that, although the specifications for positive $\delta$ and negative $\delta$ are both decreasing, the function is not decreasing on the entire real line. Since the function $\delta \le 0 \mapsto \frac {1}{\Delta} {f}^{-1}(-\delta)$ can be considered a lower bound to $\Lambda_{\Delta}(\delta)$ because we have to take into account the risk of non-execution, we can always scale the function $\delta \le 0 \mapsto \Lambda_{\Delta}(\delta)$ so that the resulting function $\Lambda_{\Delta}$ is decreasing (and strictly decreasing if we smooth the function). In general, the inequality $\Lambda'' \Lambda \le 2 \Lambda'^2$ may not be satisfied but the model can still be used although the optimal quotes $\delta^*$ may not be unique. Second, a question remains regarding the interpretation of the model when an optimal quote $\delta^*$ turns out to be negative. A possible answer, in line with the parallel made with Almgren-Chriss-like models, is to send a market order of size $\Lambda(\delta^*) = \Delta \Lambda_\Delta(\delta^*)$ (or in practice to use marketable limit orders to obtain this size).\\

\section*{Conclusion}

In this paper, we analyze optimal liquidation using limit orders. The classical literature on optimal liquidation, following Almgren-Chriss, only considers optimal scheduling and a new strand of research has recently emerged that uses either dark pools or limit orders to tackle the issue of the actual optimal way to liquidate. Our paper provides a general model for optimal liquidation with limit orders and extends both \cite{bayraktar2011liquidation} that only considers a risk-neutral framework and \cite{GLFT} that was restricted to exponential intensity functions. Our general framework also sheds new light on the important topic of negative quotes. An important improvement of our model would consist in linking the Brownian motion which drives the price and the point process modeling execution. Research in this direction has recently been made by Cartea, Jaimungal and Ricci \cite{cartea2011buy} to model market making and it may adapt to our case.

\section*{Appendix A: The multi-asset case}

This appendix is devoted to the generalization of our results to the case of a portfolio with multiple stocks. Our main result (Theorem \ref{t1}) generalizes to the multi-asset case.\\

We consider a trader who has to liquidate a portfolio made of $d$ different stocks with a quantity $q^i_0$ of stock $i$ ($i \in \lbrace 1, \ldots, d\rbrace$).

We suppose that the reference price of stock $i$ evolves as:
$$dS^i_t = \mu^i dt + \sigma^i dW^i_t,$$ with $\mathbb{V}(W^1, \ldots, W^d) = (\rho^{i,j})_{1\le i,j\le d}$ definite positive.\\

The trader under consideration continuously proposes an ask quote for each stock ($S^{i,a}_t = S^i_t + \delta^i_t$ for stock $i$), and will hence sell shares according to the rate of arrival of liquidity-taking orders at the prices he quotes.\\

The state of the portfolio is $\left(q^{1,\delta^1}, \ldots, q^{d,\delta^d}\right)$. It evolves according to the following dynamics:
$$\forall i, dq^{i,\delta^i}_t = - \Delta^i N^{i,\delta^i}_t,$$ where $N^{i,\delta^i}$ is a point process giving the number of executed orders for stock $i$, each order on stock $i$ being of size $\Delta^i$ -- we suppose as above that $\Delta^i$ is a fraction of $q^i_0$. The intensity process $(\lambda^i_t)_t$ of the point process $N^{i,\delta^i}$ is given by:

$$\lambda^i_t = \Lambda^i_{\Delta^i}(S^{i,a}_t-S^i_t)1_{q^{i,\delta^i}_{t-} > 0} = \Lambda^i_{\Delta^i}(\delta^i_t)1_{q^{i,\delta^i}_{t-} > 0},$$
where the function $\Lambda^i_{\Delta^i} : \mathbb{R} \to \mathbb{R}_+$ satisfies the same assumptions as in the single-stock case.\\

We suppose that the point processes $N^{1,\delta^1}, \ldots, N^{d,\delta^d}$ are independent.\\

The cash account $X^{\delta^1, \ldots, \delta^d}$ of the trader has then the following dynamics:\\
$$dX^{\delta^1, \ldots, \delta^d}_t = \sum_{i=1}^d (S^i_t + \delta^i_t) \Delta^i dN^{i,\delta^i}_t.$$

Eventually, the optimization problem is:

$$\sup_{(\delta^1, \ldots, \delta^d) \in \mathcal{A}^d} \mathbb{E}\left[- \exp\left(-\gamma\left(X^{\delta^1, \ldots, \delta^d}_T+\sum_{i=1}^d q^{i,\delta^i}_T (S^i_T-\ell^i(q^{i,\delta^i}_T))\right)\right) \right], $$ where the functions $\ell^1, \ldots, \ell^d$ satisfy the same assumptions as in the single-asset case.\\

This multi-asset setting deserves two remarks. First, the execution processes associated to different stocks are independent. Hence, the only difference between the multi-asset case and the single-asset cases has to do with price risk: in this multi-asset framework, optimal quotes will depend on the correlation structure between stocks. Second, the trader can only sell shares, although buying shares might sometimes reduce price risk in practice.\\

The counterpart of Proposition \ref{existuniq} and Theorem \ref{t1} in this multi-asset framework is the theorem below:

\begin{thma}[Verification theorem and optimal quotes]

There exists a unique solution $\theta_{\Delta^1, \ldots, \Delta^d}$, $C^1$ in time, of the system:
$$\forall t \in [0,T), (q^1,\ldots, q^d) \not= (0,\ldots,0),$$$$
0=\gamma \partial_t \theta_{\Delta^1, \ldots, \Delta^d}(t,q^1,\ldots,q^d) + \gamma \sum_{i=1}^d\mu^i q^i -
\frac{1}{2}  \gamma^2 \sum_{1\le i,j \le d} \rho^{i,j}\sigma^i\sigma^j q^i q^j
$$$$+ \sum_{i=1}^d 1_{q^i>0} H^i_{\Delta^i}\left(\frac{\theta_{\Delta^1, \ldots, \Delta^d}(t,q^1, \ldots, q^d) - \theta_{\Delta^1, \ldots, \Delta^d}(t,q^1, \ldots, q^i -\Delta^i, \ldots, q^d)}{\Delta^i} \right),
$$
and the conditions:
$$\theta_{\Delta^1, \ldots, \Delta^d}(T,q^1, \ldots, q^d) = -\sum_{i=1}^d \ell(q^i) q^i, \qquad \theta_{\Delta^1, \ldots, \Delta^d}(t,0,\ldots,0) = 0,$$
where
$$\forall i, H^i_{\Delta^i}(p) = \sup_{\delta^i} \Lambda^i_{\Delta^i}(\delta^i)\left( 1- e^{-\gamma \Delta^i (\delta^i - p)}\right).$$

If $\theta_{\Delta^1, \ldots, \Delta^d}$ is this function, then $$u_{\Delta^1, \ldots, \Delta^d}(t,x,q^1,\ldots,q^d,s^1,\ldots,s^d) = -\exp\left(-\gamma\left(x+\sum_{i=1}^d q^i s^i + \theta_{\Delta^1, \ldots, \Delta^d}(t,q^1, \ldots, q^d)\right)\right)$$  is the value function of the optimal control problem, and the optimal ask quotes are characterized by:
$$\forall i, (\delta^{i*}_{\Delta^i})_t = {\tilde{\delta}^{i*}_{\Delta^i}}\left(\frac{\theta_{\Delta^1, \ldots, \Delta^d}(t,q^1_t, \ldots, q^d_t) - \theta_{\Delta^1, \ldots, \Delta^d}(t,q^1_t, \ldots, q^i_t-\Delta^i, \ldots, q^d_t)}{\Delta^i}\right),$$
where ${\tilde{\delta}^{i*}_{\Delta^i}}(p)$ is uniquely characterized by:

$${\tilde{\delta}^{i*}_{\Delta^i}}(p) - \frac{1}{\gamma \Delta^i} \log\left( 1 - \gamma\Delta^i \frac{\Lambda^i_{\Delta^i} ({\tilde{\delta}^{i*}_{\Delta^i}}(p))}{{{\Lambda^i}_{\Delta^i}}' ({\tilde{\delta}^{i*}_{\Delta^i}}(p))}\right) = p$$
\end{thma}

\textbf{Proof:}\\

The proof is \emph{mutatis mutandis} the same as in the single-stock case. Existence of a local solution $t \mapsto \theta_{\Delta^1, \ldots, \Delta^d}(t,q^1,\ldots,q^d)$ comes from Cauchy-Lipschitz. To obtain existence on $[0,T]$, first notice that $\theta_{\Delta^1, \ldots, \Delta^d}(t,q^1,\ldots,q^d) +  \sum_{i=1}^d\mu^i q^i(T-t) -
\frac{1}{2}  \gamma \sum_{1\le i,j \le d} \rho^{i,j}\sigma^i\sigma^j q^i q^j(T-t)$ is a decreasing function of $t$. Now, using a comparison principle similar to Proposition \ref{comp}, we have: $$\theta_{\Delta^1, \ldots, \Delta^d}(t,q^1,\ldots,q^d) \le \sum_{i=1}^d \mu^{i+}q^i_0 (T-t) + \frac{1}{\gamma} \sum_{i=1}^d H_{\Delta^i}(0) (T-t),$$ and this bound guarantees that there is no blow up. This lead to global existence. Uniqueness follows from a comparison principle.\\
As far as verification is concerned, the proof is exactly the same as in the single-asset case.\qed\\

\section*{Appendix B: Trading on both sides}

Along with optimal liquidation, an important strand of research on optimal trading is high-frequency market making. We claimed in the introduction that many models can be used to deal with both optimal liquidation and market making. We illustrate this claim and show that our framework can easily be adapted to two-sided trading. We present the model in the single-stock case. The multi-stock case works the same but notations make the exposition cumbersome.\\

We consider a stock with a reference price following a Brownian motion with a drift:
$$dS_t = \mu dt + \sigma dW_t.$$

The trader under consideration continuously proposes a bid quote $S_t^b = S_t - \delta^b_t$ and an ask quote $S_t^a = S_t + \delta^a_t$. His inventory $q^{\delta^b,\delta^a}$ evolves according to the rate of arrival of liquidity-taking orders at the prices he quotes:
$$dq^{\delta^b,\delta^a}_t = \Delta dN^{\delta^b}_t - \Delta dN^{\delta^a}_t,$$ where $N^{\delta^b}$ and $N^{\delta^a}$ are the point processes giving the number of executed orders respectively on the bid side and on the ask side, each order being of size $\Delta$ (on both sides). The intensity processes $(\lambda^b_t)_t$ and $(\lambda^a_t)_t$ of the point processes $N^{\delta^b}$ and $N^{\delta^a}$ are given by:
$$\lambda^b_t = \Lambda_{\Delta}(S_t-S^b_t)1_{q^{\delta^b,\delta^a}_{t-} < Q} = \Lambda_{\Delta}(\delta^b_t)1_{q^{\delta^b,\delta^a}_{t-} < Q}$$ and
$$\lambda^a_t = \Lambda_{\Delta}(S^a_t-S_t)1_{q^{\delta^b,\delta^a}_{t-} > -Q} = \Lambda_{\Delta}(\delta^a_t)1_{q^{\delta^b,\delta^a}_{t-} > -Q},$$
where $\Lambda_{\Delta} : \mathbb{R} \to \mathbb{R}_+$ satisfies the same assumptions as in the case of the optimal liquidation model, and where $Q$ is a bound on the inventory. The bounds on the inventory have two roles: (i) they stand for the risk limits the traders have in practice, and (ii) they allow to write a verification theorem as in \cite{citeulike:9272221}.\\

As a consequence of his trades, the cash account  $X^{\delta^b,\delta^a}$ of the trader has the following dynamics:\\
$$dX^\delta_t = (S_t + \delta^a_t) \Delta dN^{\delta^a}_t - (S_t - \delta^b_t) \Delta dN^{\delta^b}_t.$$

Finally, the optimization problem of the high-frequency market maker is:

$$\sup_{\delta^b,\delta^a \in \mathcal{A}} \mathbb{E}\left[- \exp\left(-\gamma\left(X^{\delta^b,\delta^a}_T+q^{\delta^b,\delta^a}_T S_T - |q^{\delta^b,\delta^a}_T|\ell(|q^{\delta^b,\delta^a}_T|)\right)\right) \right],$$ where $\ell$ satisfies the same assumptions as in the case of the optimal liquidation model.\\

The counterpart of Proposition \ref{existuniq} and Theorem \ref{t1} in this market model framework is the theorem below:

\begin{thmb}[Verification theorem and optimal quotes]

There exists a unique solution $\theta_{\Delta}$, $C^1$ in time, of the system:
$$\forall t \in [0,T), \forall q \in \{ -Q, \ldots, -\Delta, 0, \Delta, \ldots, Q \},\quad 0=\gamma \partial_t \theta_{\Delta}(t,q) + \gamma \mu q -
\frac{1}{2}  \gamma^2 \sigma^2 q^2
$$$$+ 1_{q < Q}H_{\Delta}\left(\frac{\theta_{\Delta}(t,q) - \theta_{\Delta}(t,q+\Delta)}{\Delta} \right) + 1_{q > -Q}H_{\Delta}\left(\frac{\theta_{\Delta}(t,q) - \theta_{\Delta}(t,q-\Delta)}{\Delta} \right)
$$
and the terminal condition:
$$\theta_{\Delta}(T,q) = - \ell(|q|)|q|,$$
where $H_{\Delta}(p) = \sup_{\delta} \Lambda_{\Delta}(\delta)\left( 1- e^{-\gamma \Delta (\delta - p)}\right).$\\

If $\theta_{\Delta}$ is this function, then $$u_{\Delta}(t,x,q,s) = -\exp\left(-\gamma\left(x+qs + \theta_{\Delta}(t,q)\right)\right)$$  is the value function of the optimal control problem, and the optimal bid and ask quotes are characterized by:

$$(\delta_{\Delta}^{b*})_t = \tilde{\delta}_{\Delta}^*\left(\frac{\theta_\Delta(t,q_t) - \theta_\Delta(t,q_t+\Delta)}{\Delta}\right),$$
$$(\delta_{\Delta}^{a*})_t = \tilde{\delta}_{\Delta}^*\left(\frac{\theta_\Delta(t,q_t) - \theta_\Delta(t,q_t-\Delta)}{\Delta}\right),$$
where $\tilde{\delta}_{\Delta}^*(p)$ is uniquely characterized by the equation $\left(E_{\delta^*_\Delta}\right)$ of Lemma \ref{l1}.
\end{thmb}

\textbf{Proof:}\\

The proof is close to the proof in the one-sided case. Existence of a local solution $t \mapsto \left(\theta_{\Delta}(t,q)\right)_{q \in \{ -Q, \ldots, -\Delta, 0, \Delta, \ldots, Q \}}$ comes from Cauchy-Lipschitz. To obtain existence on $[0,T]$, first notice that $\theta_{\Delta}(t,q) +  \mu q(T-t) - \frac{1}{2}  \gamma \sigma^2 q^2(T-t)$ is a decreasing function of $t$. Then, using a comparison principle, we have: $$\theta_{\Delta}(t,q) \le |\mu|Q (T-t) + \frac{2}{\gamma} H_{\Delta}(0) (T-t),$$ and this bound guarantees that there is no blow up, hence global existence. Uniqueness follows from a comparison principle.\\
As far as verification is concerned, the proof similar to the proof of Theorem \ref{t1}.\qed\\

\bibliographystyle{plain}

\end{document}